\documentclass[10pt]{article}

\usepackage{cite}
\usepackage{tikz}
\usepackage[permil]{overpic}
\usepackage{subfig}
\usepackage{array}

\newcommand{\vpp}{\textsc{Voro++}}

\makeatletter
\def\@fnsymbol#1{\ensuremath{\ifcase#1 \or 1\or 2\or 3\or 4\or 5\or 6\or 7
\or \ddagger\ddagger \else\@ctrerr\fi}}
\makeatother

\begin{document}

\title{Characterizing structural features \\of two-dimensional particle systems \\through Voronoi topology}
\author{
Emanuel A. Lazar\thanks{Department of Mathematics, Bar Ilan University, Ramat Gan 5290002, Israel}\,\,, 
Jiayin Lu\thanks{Department of Mathematics, University of California, Los Angeles, CA 90095, USA}\,\,$^,$\thanks{Department of Mathematics, University of Wisconsin--Madison, Madison, WI 53711, USA}\,\,,\\ 
Chris H. Rycroft\footnotemark[3]\,$\,^,$\thanks{Mathematics Group, Lawrence Berkeley Laboratory, Berkeley, CA 94720, USA}\,\,, 
Deborah Schwarcz\footnotemark[1]\,$\,^,$\thanks{Soreq Nuclear Research Center, Yavne 81800, Israel}}
\date{\today}
\maketitle

\begin{abstract}
This paper introduces a new approach toward characterizing local structural features of 
two-dimensional particle systems. The approach can accurately identify and characterize 
defects in high-temperature crystals, distinguish a wide range of nominally disordered 
systems, and robustly describe complex structures such as grain boundaries. This paper 
also introduces two-dimensional functionality into the open-source software program 
\textit{VoroTop} which automates this analysis. This software package is built on a 
recently-introduced multithreaded version of \vpp, enabling the analysis of systems with 
billions of particles on high-performance computer architectures.
\end{abstract}

\section{Introduction}

Many two-dimensional physical systems can be studied as large sets of point-like particles, 
and the arrangement of these particles in space often determines many of these systems' 
chemical, electronic, and mechanical properties \cite{wood13, lin2016defect, wu2017spectroscopic, zhang2018defect, tanjeem21, lobmeyer2022grain}.
\setlength{\fboxsep}{0pt}
\begin{figure}
\begin{center}
\fbox{\begin{overpic}[width=0.32\columnwidth]{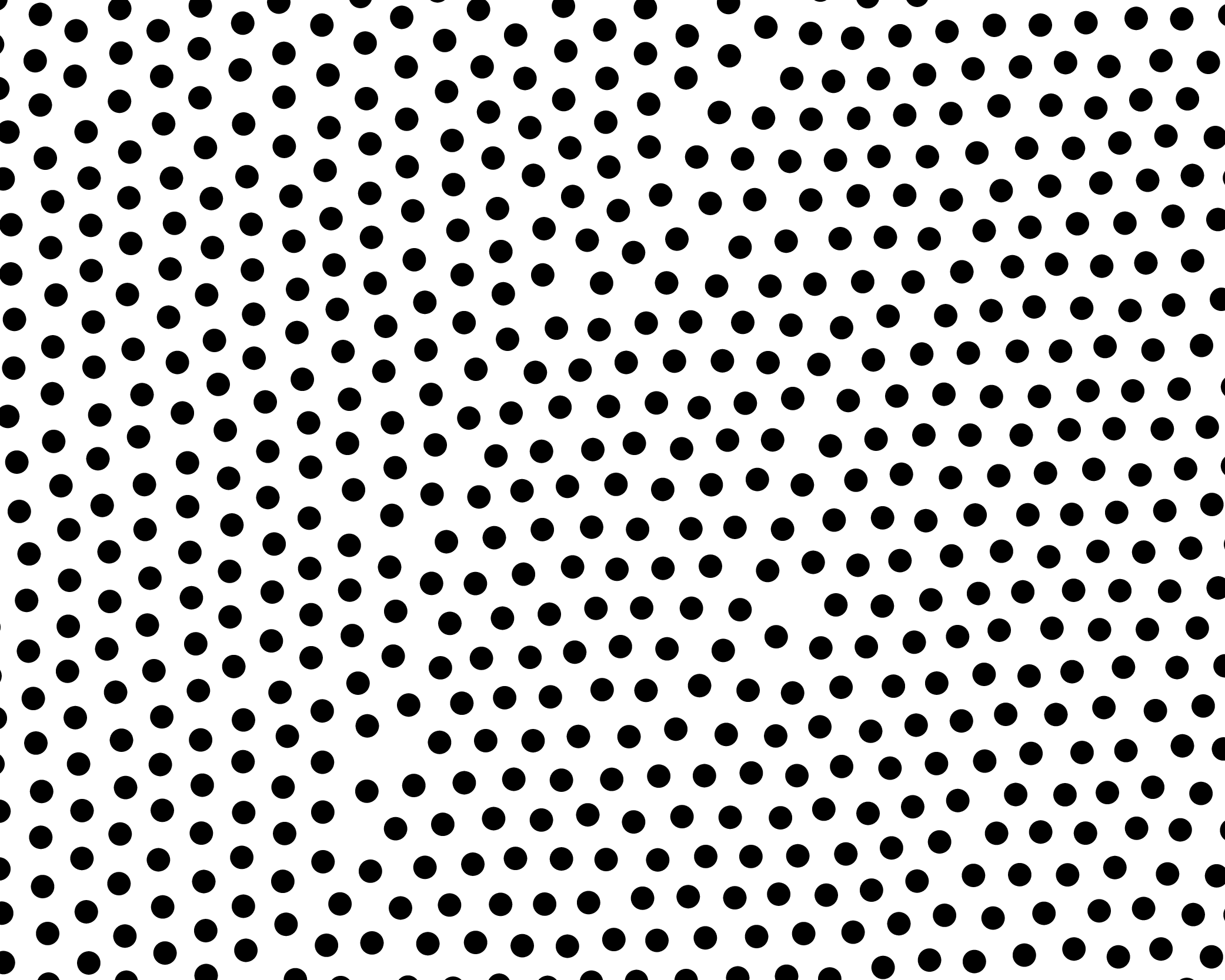}
\put (0,690.) {\setlength{\fboxsep}{1pt} {\fcolorbox{black}{white}{(a)}}}
\end{overpic}} \hfill
\fbox{\begin{overpic}[width=0.32\columnwidth]{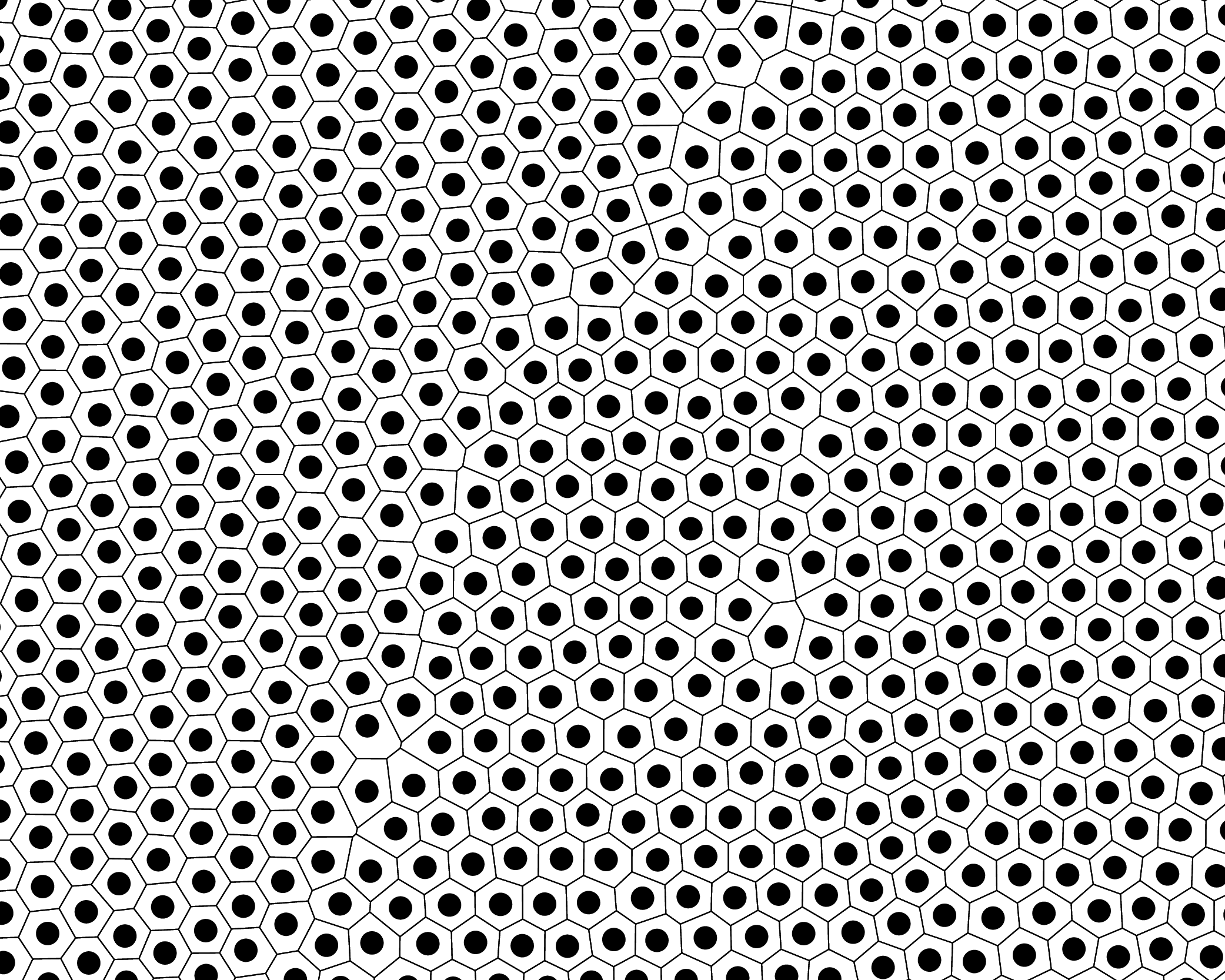}
\put (0,690.) {\setlength{\fboxsep}{1pt} {\fcolorbox{black}{white}{(b)}}}
\end{overpic}} \hfill
\fbox{\begin{overpic}[width=0.32\columnwidth]{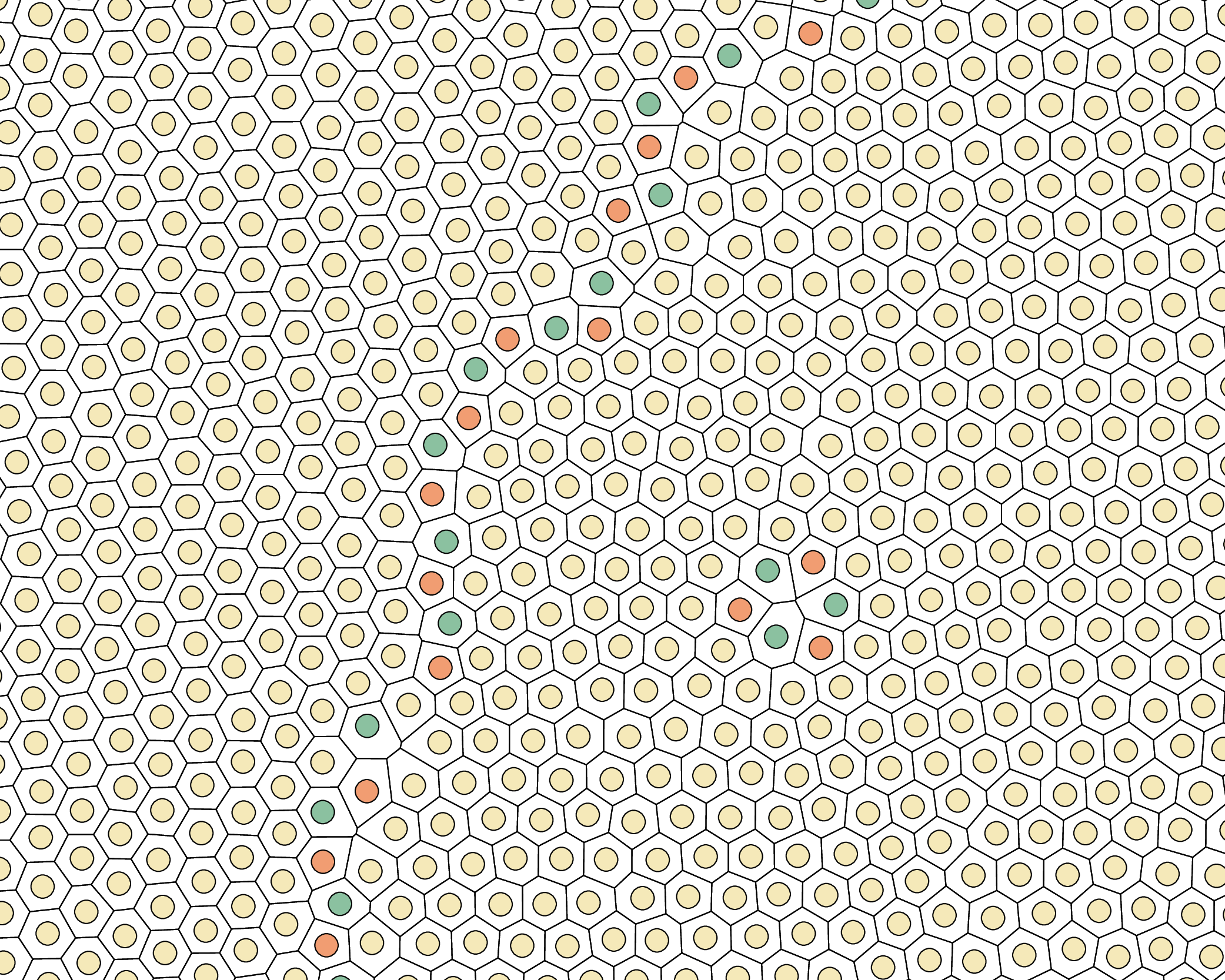}
\put (0,690.) {\setlength{\fboxsep}{1pt} {\fcolorbox{black}{white}{(c)}}}
\end{overpic}}
\end{center}
\caption{Two adjacent crystals separated by a grain boundary and containing a vacancy. 
The system was created using molecular dynamics through the cooling of a Lennard-Jones liquid: (a) particles, (b) particles and their Voronoi cells, (c) particles colored by the number of edges of their Voronoi cells.}
\label{fig:two-dimensional}
\end{figure}
It is therefore important to have available precise, robust, and efficient tools that can 
automatically identify structural objects such as crystals and defects in large atomistic data 
sets. Figure \ref{fig:two-dimensional} illustrates a pair of adjacent crystals, separated by a grain boundary 
and containing a vacancy. Although the rough 
contours of these defects can be observed visually, identifying them precisely enough for 
automated, quantitative analysis is challenging.

Recent decades have witnessed the development of powerful tools to automate the 
identification and analysis of structural objects in large atomistic data sets 
\cite{stukowski2012structure, Tomaso_review}. Many of these approaches describe arrangements of particles by quantifying their similarity to an ideal reference arrangement with respect to some property. For example, some methods count 
the number of particles in a fixed range from each central particle \cite{hoekstra2003flow}, 
quantify the variation in distances to neighbors \cite{PhysRevLett_Ultrafast_uenching}, or 
else the variation in angles between neighboring particles \cite{halperin1978theory, 
hamanaka2006transitions}. Other methods quantify the degree to which the neighborhood is 
centrosymmetric \cite{Kelchner_centrosymmetry}, a defining feature of lattice crystals.

These approaches typically require carefully-chosen cutoffs for analyzing different kinds of 
systems. Moreover, such approaches are typically ineffective for characterizing particular 
kinds of defects. Perhaps most significantly, although these methods are typically well-suited 
for studying systems at low temperatures, they often perform poorly when applied to 
systems at high temperatures, or otherwise strongly perturbed from their ground state 
\cite{stukowski2012structure}. Topological approaches tend to be more effective, due to the 
method in which they segment data in a high-dimensional configuration space, instead of in 
an image of that space under a continuous mapping \cite{landweber2016fiber, 
lazar2015topological}.
Numerous methods based on machine learning have also been introduced in recent years 
\cite{reinhart2017machine, chung2022data, lafourcade2023robust}, though these methods 
do not characterize crystalline structure directly.

This paper introduces a new, simple approach for classifying structure in two-dimensional 
particle systems. This approach is based on Voronoi topology and thus naturally ignores 
small fluctuations in particle positions associated with thermal vibrations and small strains, 
without the need for quenching, temporal averaging, or arbitrary order-parameter cutoffs. 
The method is further useful for studying both ordered and nominally disordered systems. 
Many ideas suggested here can be considered as adaptations and extensions of ideas 
introduced and developed previously for three-dimensional systems 
\cite{lazar2015topological, lazar2017vorotop, lazar2018}.

In addition to developing a new approach towards characterizing structure in 
two-dimensional particle systems, this paper also introduces two-dimensional functionality 
into the open-source command-line program called \textit{VoroTop} to automate this 
analysis. The latest version of \textit{VoroTop} is designed to utilize a recently-introduced, 
multithreaded version of the \vpp\, library \cite{2009rycroft, lu2023extension} for computing 
Voronoi cells, enabling the study of large systems with billions of particles.

This paper is organized as follows. Section \ref{sec:basics} describes the basics of Voronoi 
cells, and explains how their topology can be used to characterize and analyze structure in 
two-dimensional particle systems. Section \ref{sec:software} describes the two-dimensional 
\textit{VoroTop} functionality and its core functions and features. Section 
\ref{sec:applications} illustrates several example applications, including the identification of 
defects in crystals, the characterization of order in disordered systems, and the analysis of 
grain boundaries, including chiral features, in non-ideal systems.

\section{Voronoi topology}
\label{sec:basics}

\subsection{Voronoi cells and their shapes}
In a system of discrete particles, the \textit{Voronoi cell} of each particle is the region of 
space closer to it than to any other particle \cite{voronoi1908nouvelles, okabe2009spatial, 
aurenhammer1991voronoi}. Figure \ref{fig:two-dimensional} illustrates a bicrystal and 
Voronoi cells of some of the particles.
Geometric and topological features of a Voronoi cell can be used to characterize features of 
local ordering in the vicinity of each particle \cite{lazar2022voronoi}. For example, particles 
can be defined as neighbors if they share a Voronoi edge, so that the number of Voronoi cell 
edges gives a count of neighbors. In a defect-free hexagonal crystal, even at temperatures 
above zero, the Voronoi cell of each particle has six edges. In crystals containing defects the 
Voronoi cells of some particles will have other numbers of edges. This can be vividly 
observed in Figure \ref{fig:two-dimensional}(c), in which many Voronoi cells have five and 
seven edges.

\begin{figure}[b]
\begin{center}
\fbox{\begin{overpic}[height=0.2\columnwidth]{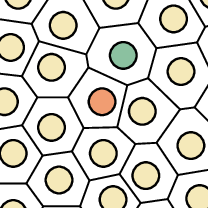}
\put (-10,830.) {\setlength{\fboxsep}{1pt} {\fcolorbox{black}{white}{(a)}}}
\end{overpic}} \hspace{1mm}
\fbox{\begin{overpic}[height=0.2\columnwidth]{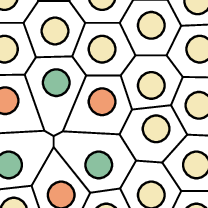}
\put (-10,830.) {\setlength{\fboxsep}{1pt} {\fcolorbox{black}{white}{(b)}}}
\end{overpic}} \hspace{1mm}
\fbox{\begin{overpic}[height=0.2\columnwidth]{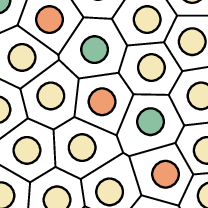}
\put (-10,830.) {\setlength{\fboxsep}{1pt} {\fcolorbox{black}{white}{(c)}}}
\end{overpic}} \hspace{2mm}
\includegraphics[height=0.2\columnwidth]{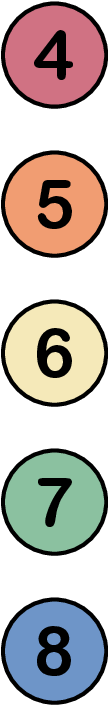}
\end{center}
\caption{(a-c) Three central particles, each with five neighbors, associated to distinct structural defects.  These structural differences are reflected in differences in the numbers of edges of neighboring Voronoi cells.  Particle colors indicate numbers of Voronoi edges or neighbors.}
\label{fig:three-fives}
\end{figure}

The number of edges of a Voronoi cell, however, is a rather coarse description of local 
structure in particle arrangements. As can be seen in Figure \ref{fig:two-dimensional}(c), 
five- and seven-sided Voronoi cells are associated with grain boundaries and vacancies, 
and so the number of edges alone provides only modest structural information.
A more refined description of local arrangements of particles can provide a more nuanced, 
and useful, description. In particular, we characterize each particle according to its number 
of edges and the number of edges of its neighbors, ordered sequentially. Figure 
\ref{fig:three-fives} illustrates three central particles and their Voronoi cells. Although each of 
the three central particles have Voronoi cells with five neighbors, these neighbors have 
different numbers of neighbors themselves, reflecting distinct local orderings. In Figure 
\ref{fig:three-fives}(a), all neighbors have six neighbors except for one that has seven; this 
particle is associated to a dislocation. The two particles in Figures \ref{fig:three-fives}(b) and 
(c) each have three six-sided neighbors and two five-sided ones. They are structurally 
distinct, however, in that the two five-sided neighbors are adjacent in Figures 
\ref{fig:three-fives}(b) but not in (c); one belongs to a vacancy while the other belongs to a 
grain boundary.

The number of edges of the Voronoi cells of a particle and its neighbors thus provides a 
simple description of particle arrangements that can distinguish particles associated with 
different kinds of defects. Since topological features of Voronoi cells do not change under 
rotations, translations, or rescalings, this description is consistent with the intuition that such 
transformations do not impact structurally significant features of a system. Furthermore, 
since topological features of Voronoi cells do not generally change under small perturbations 
of particle coordinates, characterization will typically be insensitive to small measurement 
errors. 
Finally, even in special cases in which small perturbations will result in discrete shifts 
in topology, those shifts can be completely understood and the resulting topologies are fully 
described in statistically precise terms.
All of this might be contrasted with approaches that rely on geometric features, such as 
Voronoi cell areas or perimeters.  These quantities typically change under perturbations, 
and methods constructed based on them consequentially require choosing cutoffs, often 
somewhat arbitrary, for classification.

\subsection{Canonical representations}
\label{sec:canonical}

We thus use the term \textit{Voronoi topology} of a particle to refer to the number of edges of 
its Voronoi cell and those of its neighbors, ordered sequentially. For each particle whose 
Voronoi cell has $n$ edges, this information is represented by an ordered list of $n+1$ 
numbers. The first counts the neighbors of the central particle, equivalently the number of 
edges of its Voronoi cell; subsequent numbers count the numbers of neighbors of 
neighboring particles. This description might be considered a two-dimensional analogue of 
the Weinberg codes considered elsewhere for characterizing three-dimensional polyhedra 
\cite{weinberg1965plane, 1966weinberg1, 2012lazar}.
We note that the first element of the $p$-vector is currently redundant, since the information 
can be inferred from the length of the vector.  It is included to facilitate future generalizations, 
such as to multicomponent systems.

As an example, we consider the arrangement of particles in Figure \ref{fig:three-fives}(c). 
The central particle has five neighbors. If we enumerate the number of edges of its 
neighbors in counterclockwise order beginning with the particle to its left, we arrive at the 
sequence $(5,6,6,7,6,7)$. If we had instead begun with the neighbor above, we would arrive 
at the sequence $(5,7,6,6,7,6)$; beginning with other neighbors can result in other 
sequences. Since all of these sequences describe the same structural information, we 
choose the lexicographically first one as the canonical representation of the Voronoi 
topology, and use the term \textit{$p$-vector} to denote it. We say that two arrangements of 
particles have the same Voronoi topology if their $p$-vectors are identical.

A subtle issue arises when considering orientation. In some cases, had we enumerated the 
number of edges of neighbors in a clockwise manner instead of in a counter-clockwise 
manner, we would arrive at a different set of sequences and a different \textit{$p$-vector}, 
indicating a chirality, or handedness, of the arrangement. In other cases, the two orientations 
generate the same sets of sequences, indicating a mirror symmetry in the arrangement.

We thus establish the following convention. We calculate the sequences associated with 
both orientations and choose the lexicographically first one among all sequences as the 
\textit{canonical} $p$-vector; we also store information about the chirality of the arrangement 
for further analysis. In particular, if the sequences for the two orientations are identical, then 
the arrangement is non-chiral. Otherwise, if the lexicographically-first sequence is in the 
counterclockwise list, we say that it is left-handed, and if the lexicographically-first sequence 
is in the clockwise list, we say that it is right-handed. Section \ref{sec:chiral} illustrates an 
example system in which Voronoi topology is able to detect chirality of a grain boundary.

\begin{figure}
\centering
\hfill
\hfill
\fbox{\begin{overpic}[width=0.25\columnwidth]{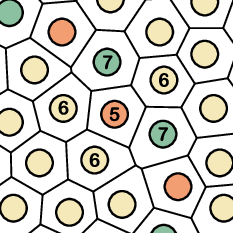}
\put (0.,840) {\setlength{\fboxsep}{1pt} {\large \fcolorbox{black}{white}{(a)}}}
\end{overpic}}
\hfill
\fbox{\begin{overpic}[width=0.25\columnwidth]{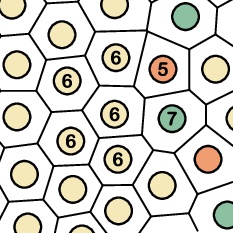}
\put (0.,840) {\setlength{\fboxsep}{1pt} {\large \fcolorbox{black}{white}{(b)}}}
\end{overpic}}
\hfill
\hfill
\hfill
\caption{(a) In this arrangement of particles, $p$-vector descriptions of neighboring edges 
are identical whether we list neighbors sequentially in clockwise or counterclockwise fashion. 
(b) In this arrangement, $p$-vector descriptions will differ depending on 
whether we list neighbors in a clockwise or counterclockwise manner.}
\label{fig:twosets}
\end{figure}

To illustrate the procedure for constructing a canonical $p$-vector, Table 
\ref{table:sequences} lists the integer sequences that describe the Voronoi topology of the 
arrangements in Figures \ref{fig:twosets}(a) and (b) for each of two orientations, sorted in 
lexicographical order. The neighborhood illustrated in Figure \ref{fig:twosets}(a) is non-chiral 
and the two sets of sequences are identical. In contrast, the arrangement shown in Figure 
\ref{fig:twosets}(b) has a handedness, and the two sets of sequences are different. Since the 
lexicographically first sequence of numbers results from a counterclockwise enumeration of 
neighbor edges, this arrangement is considered left-handed.

\begin{table}
\hfill
\begin{tabular}{|| l | l ||}
 \hline
\multicolumn{1}{ || c }{Orientation 1} & \multicolumn{1}{ | c || }{Orientation 2} \\
 \hline\hline
$(5, 6, 6,7, 6, 7)$ & $(5, 6, 6,7, 6, 7)$ \\
$(5, 6, 7, 6, 6,7)$ & $(5, 6, 7, 6, 6,7)$ \\
$(5, 6, 7, 6, 7,6)$ & $(5, 6, 7, 6, 7,6)$ \\
$(5, 7, 6, 6,7,6)$ & $(5, 7, 6, 6,7,6)$ \\
$(5, 7, 6, 7, 6, 6)$& $(5, 7, 6, 7, 6, 6)$ \\ &\\
\hline
\end{tabular}
\hfill
\begin{tabular}{|| l | l ||}
 \hline
\multicolumn{1}{ || c }{Orientation 1} & \multicolumn{1}{ | c || }{Orientation 2} \\
 \hline\hline
$(6, 5, 6, 6, 6, 6, 7)$ & $(6,5, 7, 6, 6, 6, 6)$ \\
$(6, 6, 6, 6, 6, 7,5)$ & $(6, 6, 5,7,6,6,6)$  \\
$(6, 6, 6, 6, 7,5,6)$ & $(6, 6, 6, 5,7,6,6)$ \\
$(6, 6, 6, 7,5,6,6)$ & $(6, 6, 6, 6, 5,7,6)$ \\
$(6, 6, 7,5,6,6,6)$ & $(6, 6, 6, 6, 6, 5,7)$\\
$(6, 7, 5, 6, 6, 6,6)$ & $(6, 7, 6, 6, 6, 6,5)$ \\
\hline
\end{tabular}
\hfill
\caption{Lists of integer sequences associated with particle arrangements illustrated in 
Figures \ref{fig:twosets}(a) and (b). The sequences associated with the clockwise and 
counterclockwise orientations for Figure \ref{fig:twosets}(a) are identical, indicating a mirror 
symmetry and lack of handedness. In contrast, the sequences for the two orientations are 
different for the arrangement in Figure \ref{fig:twosets}(b), indicating a handedness. Since 
the lexicographically lowest sequence is in the counterclockwise list, we call the 
arrangement left-handed.}
\label{table:sequences}
\end{table}

\subsection{Perturbation analysis}
\label{sec:perturbations}

Small perturbations of particle coordinates do not always change their Voronoi topologies. 
For example, in the bicrystal illustrated in Figure \ref{fig:two-dimensional}, the Voronoi 
cells of most particles are six-sided, even though the particles are slightly perturbed from 
ideal lattice positions as a result of thermal vibrations and small internal strains. Similarly, the 
Voronoi topologies of particles associated with the grain boundary and vacancy are also 
stable under small perturbations of particle coordinates.

\begin{figure}[b]
\begin{center}
\begin{overpic}[width=0.5\columnwidth]{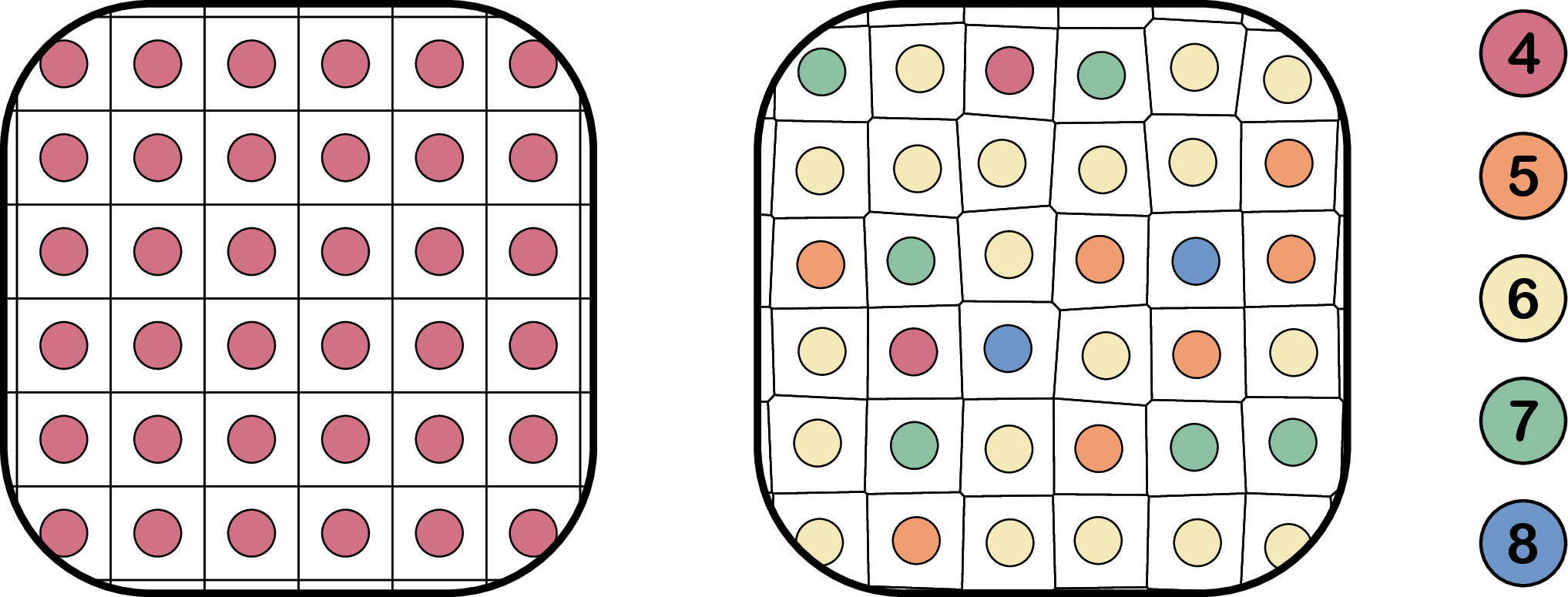}
\put (-80,350) {(a)} 
\put (400,350) {(b)}
\end{overpic}
\vspace{-3mm}
\end{center}
\caption{Particles colored according to the number of edges of their Voronoi cells in (a) an 
unperturbed square lattice and (b) a perturbed square lattice.}
\label{fig:square_perturbed}
\end{figure}
In some arrangements, however, especially those associated with perfect crystals and 
idealized defects, Voronoi topology can change under small perturbations such as those 
associated with thermal noise, small strains, or measurement error. As an example, Figure 
\ref{fig:square_perturbed} illustrates a square lattice; the $p$-vector of every particle is 
$(4,4,4,4,4)$. Under small perturbations, however, corners of Voronoi cells can resolve into 
edges and the resulting Voronoi cells can have between 4 and 8 edges each, resulting in 
many different $p$-vectors.
Similarly, particles associated to the ``ideal'' vacancy arrangement illustrated in Figure 
\ref{fig:vacancy}(c) all have the $p$-vector $(5,5,5,6,6,6)$. Small perturbations of the particle 
coordinates, however, can result in different Voronoi topologies of the associated particles, 
as illustrated in Figures \ref{fig:vacancy}(d-f).

We therefore consider the possibility that a given structural object such as a crystal or defect 
can be associated with multiple Voronoi topologies, equivalently $p$-vectors, under 
infinitesimal perturbations. We use the term {\it family} to denote a set of Voronoi topologies 
that can be obtained from an ideal structure through infinitesimal perturbations of particle 
coordinates. Particles whose Voronoi topologies belong to a family of crystalline Voronoi 
topologies are classified as belonging to a bulk crystal, while those whose topologies belong 
to a family of defect topologies are classified as belonging to defects.

Families of Voronoi topologies associated with particular crystals and defects can sometimes 
be determined analytically by consideration of possible resolutions of individual unstable 
corners \cite{lazar2015topological, lazar2018, leipold2016statistical, lazar2020voronoi}. 
However, this approach is often complicated by analytical and computational challenges.

\overfullrule=0pt
\setlength{\fboxsep}{0pt}
\setlength{\tabcolsep}{0.2em}
\begin{figure}
\centering
\fbox{\begin{overpic}[height=0.15\linewidth]{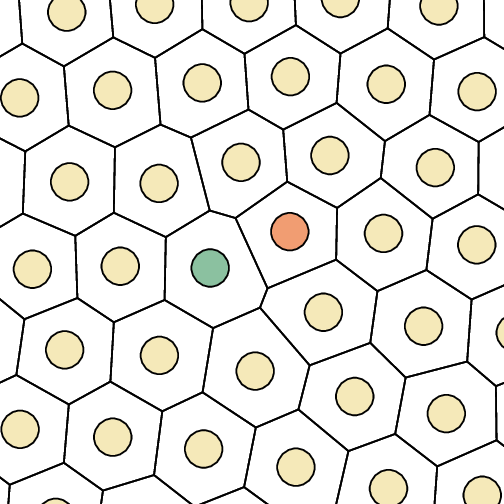}
\put (-20.,830) {\setlength{\fboxsep}{1pt} {\scriptsize \fcolorbox{black}{white}{(a)}}}
\end{overpic}}
\fbox{\begin{overpic}[height=0.15\linewidth]{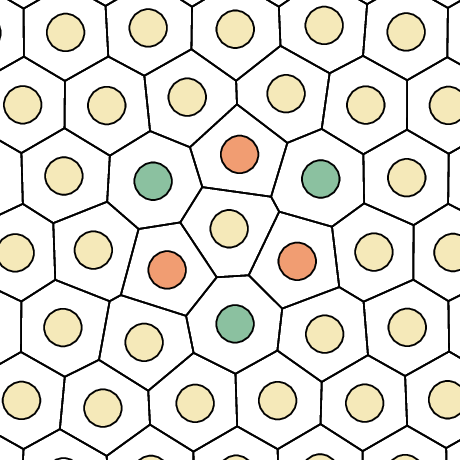}
\put (-20.,830) {\setlength{\fboxsep}{1pt} {\footnotesize \fcolorbox{black}{white}{(b)}}}
\end{overpic}}
\hfill
\fbox{\begin{overpic}[height=0.15\linewidth]{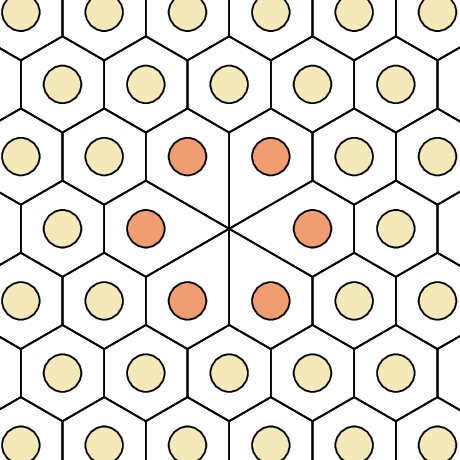}
\put (-20.,830) {\setlength{\fboxsep}{1pt} {\footnotesize \fcolorbox{black}{white}{(c)}}}
\end{overpic}}
\fbox{\begin{overpic}[trim={2.cm 2.cm 2.cm 2.cm},clip,height=0.15\linewidth]{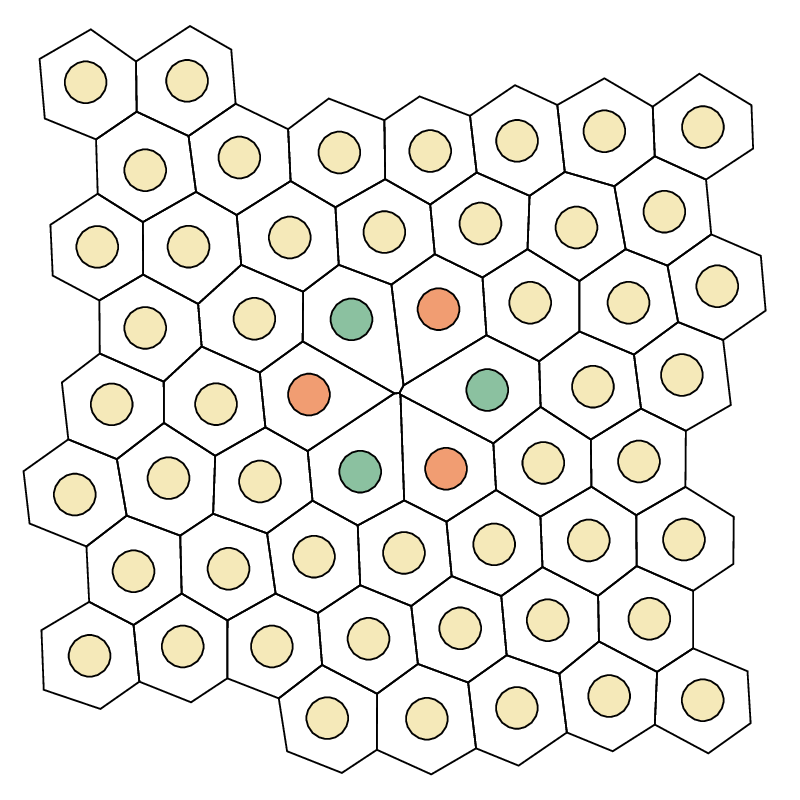}
\put (-20.,830) {\setlength{\fboxsep}{1pt} {\footnotesize \fcolorbox{black}{white}{(d)}}}
\end{overpic}}
\fbox{\begin{overpic}[height=0.15\linewidth]{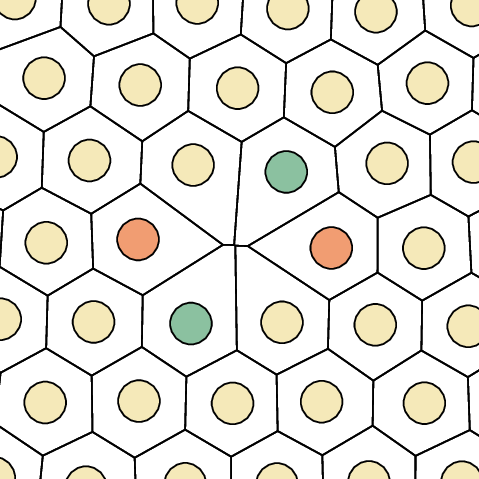}
\put (-20.,830) {\setlength{\fboxsep}{1pt} {\footnotesize \fcolorbox{black}{white}{(e)}}}\end{overpic}}
\fbox{\begin{overpic}[trim={2.4cm 2.4cm 2.4cm 2.4cm},clip,height=0.15\linewidth]{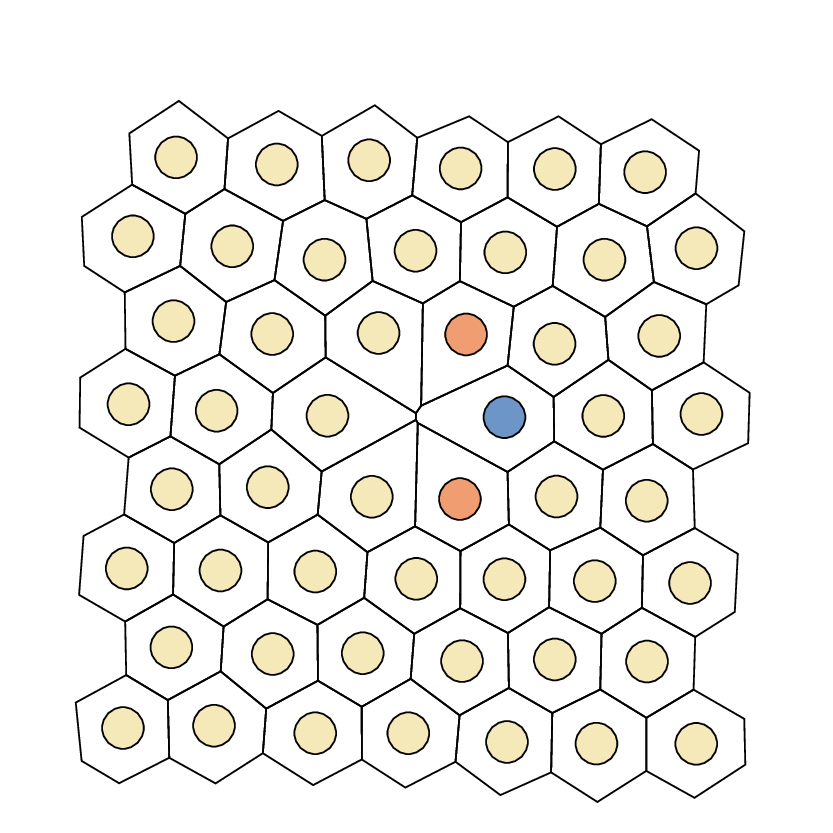}
\put (-20.,830) {\setlength{\fboxsep}{1pt} {\footnotesize \fcolorbox{black}{white}{(f)}}}
\end{overpic}}
\hfill
\includegraphics[height=0.15\columnwidth]{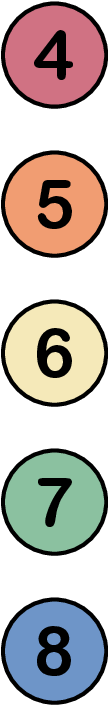}
\caption{(a) An isolated dislocation and (b) interstitial with stable Voronoi topologies; (c) an unstable vacancy resolves under perturbations as either (d), (e), or (f).}
\label{fig:vacancy}
\end{figure}

\subsection{Cluster analysis}
\label{sec:clusteranalysis}

The proposed approach characterizes the local ordering of individual particles. Analysis of 
contiguous groups of particles with particular local structural classification can be 
subsequently used to identify and analyze larger-scale structural objects.
A defect-free hexagonal crystal, for example, can be defined as a set of contiguous particles 
all of which have $p$-vector $(6,6,6,6,6,6,6)$. Likewise, interstitial defects can be identified 
with a particle with $p$-vector $(6,5,7,5,7,5,7)$ surrounded by six neighboring particles with 
alternating $p$-vectors $(5,6,6,7,6,7)$ and $(7,5,6,5,6,6,6,6)$, as illustrated in Figure 
\ref{fig:vacancy}(b). This topological approach to characterizing defects is general in that it 
can be used to characterize and subsequently identify different kinds of crystals and defects. 
At the same time, this method is robust in that small perturbations of particle coordinates do 
not generally affect this structural classification.

In a similar manner we can also identify contiguous regions of unspecified non-crystalline 
order inside a crystalline system. In studying mechanisms such as melting, this approach 
provides a well-defined, robust method for characterizing different parts of a system as 
crystalline or not, facilitating quantitative analysis of growth and degradation of phases within 
a larger matrix such as those that occur under conditions suitable for phase transformations.

\subsection{Indeterminate types and their resolutions}
\label{sec:indeterminate}
A complication that arises in enumerating families of Voronoi topologies is the possibility that 
a topology belongs to multiple structural families; we call such topologies 
\textit{indeterminate}.
As a concrete example, the Voronoi topology denoted by the $p$-vector $(6,6,6,6,6,6,6)$ 
and that appears in hexagonal crystals also appears in perturbations of a square lattice 
\cite{leipold2016statistical,lazar2020voronoi}. Likewise, the $p$-vector $(5,6,6,6,6,7)$ is 
associated with both an isolated dislocation, as illustrated in Figure \ref{fig:vacancy}(a), as 
well as a vacancy, as illustrated in Figure \ref{fig:vacancy}(e). To complicate the matter 
further, this topology can also appear at the end of a grain boundary. 

These indeterminacies can be resolved in several ways. One approach involves 
consideration of probabilities of the indeterminate topologies appearing in various systems. 
For example, in a defect-free hexagonal crystal, all particles have the Voronoi topology given 
by the $p$-vector $(6,6,6,6,6,6,6)$. In contrast, this topology appears in the perturbed 
square lattice and ideal gas with extremely small probabilities \cite{lazar2020voronoi}. If we 
find such an arrangement in a general system, we might conclude that it more likely belongs 
to an hexagonal crystal than to a square one or to an ideal gas.
This approach, however, is unsatisfactory since it suggests that every particle characterized 
by the $p$-vector $(6,6,6,6,6,6,6)$ be classified as having hexagonal local structure, 
including those that appear in a square crystal or ideal gas. Large square lattice crystals with 
arbitrarily small random perturbations would then typically include particles classified as 
defects.

A second approach involves randomly perturbing particle positions and recomputing their 
topologies. We can repeat this process several times and classify the local structure 
according to whether the majority of resolutions result in a determinate topology of one kind 
or another. Such analysis was suggested in a paper describing an earlier version of 
\textit{VoroTop} \cite{lazar2017vorotop}. This approach, however, requires computing 
Voronoi cells multiple times per particle. Moreover, it also requires a default-case analysis so 
that $(6,6,6,6,6,6,6)$ would be classified as hexagonal if no perturbations resulted in a 
determinate square lattice topology.

We thus suggest a third approach that builds on the analysis described in Section 
\ref{sec:clusteranalysis}.  In particular, after classifying structure types of individual 
particles, we construct clusters of particles that are identified as non-crystalline and whose 
Voronoi cells are contiguous.  To resolve indeterminate types, we then consider the 
Voronoi topologies of the particles that constitute the cluster.  A dislocation, for example, 
consists of an adjacent pair of particles, one with Voronoi topology described by the 
$p$-vector $(5,6,6,6,6,7)$, and one by $(7,5,6,6,6,6,6,6)$.  Although such topologies can 
also appear individually at the ends of a grain boundary, knowing that they belong to a 
defect cluster with only one of each is sufficient to resolve them as constituting a dislocation.  
An example demonstrating this kind of analysis can be found in Section \ref{sec:sub:identifying}.
This approach might be contrasted with a 
mean-field approach developed and previously applied to disordered systems 
\cite{yoon2018topological}.

\section{\textit{VoroTop} software}
\label{sec:software}

The open-source \textit{VoroTop} software package was developed to automate the analysis 
of structural features in particle systems using Voronoi topology \cite{lazar2017vorotop}. The 
program was initially designed for three-dimensional systems. We now describe extensions 
to automate analysis of two-dimensional systems.

\subsection{Language, license, and availability}
The \textit{VoroTop} software package is written in \texttt{C++11} and is compatible with all major operating 
systems.  \textit{VoroTop} is released under an OpenSource BSD 3-Clause license, which permits 
redistribution and use of source and binaries, with or without modification, to both academic 
and for-profit groups. \textit{VoroTop} is available online in a Git repository at 
\texttt{https://gitlab.com/mLazar/VoroTop/}.

\subsection{Performance, optimization, and runtime}
The latest version of \textit{VoroTop} is built using a new version of \vpp{} 
\cite{2009rycroft,lu2023extension} that incorporates multithreading with 
OpenMP~\cite{dagum98}. The \vpp{} library computes Voronoi cells individually, and the 
total computation time scales approximately linearly for typical, dense particle 
arrangements. Since each Voronoi cell can be computed independently of others, the 
multithreaded version has near-optimal parallel efficiency, in both two and three 
dimensions~\cite{lu2023extension}. The computation of the $p$-vectors in \textit{VoroTop} 
is also multithreaded, and typically takes a constant amount of additional work per 
particle. Running with a single thread on an Intel Xeon Gold 6240 CPU running at 
2.60\,GHz, \textit{VoroTop} can currently compute about 160,000 Voronoi cells and 
$p$-vectors per second, or roughly ten million particles per minute.

\subsection{Filters}
\label{sec:filters}

We use the term \textit{filter} to refer to a list of one or more families of Voronoi topologies used by \textit{VoroTop} 
to identify crystalline and defect structure. As a simple example, a filter can enumerate only the 
unique $p$-vector $(6,6,6,6,6,6,6)$ associated with the ideal hexagonal lattice, or else also 
list families associated with defects such as dislocations, vacancies, and grain boundaries.

\textbf{File format.} Filter files are divided into three parts. 
The first part consists of optional comments about the filter, such as its source, statistical analysis, or other notes; all lines that begin with a `\texttt{\#}' are treated as comments. 
Lines in the second part begin with a `\texttt{*}' and 
specify user-defined structure types. Each such line, after the `\texttt{*}', includes an index and a name for 
the structure type. Indices of structure types are listed in increasing order and 
begin with 1. 
The third part consists of lines that record Voronoi cell topologies, represented by $p$-vectors, and their associated structure types. Each line begins with a structure type index and an associated 
$p$-vector, as described above. Topologies listed as 
belonging to multiple structure types are indeterminate.
Filter files for several common structure types, included those considered in this paper, can be found at \texttt{www.vorotop.org}.

\textit{VoroTop} begins by reading in information about a system represented in the LAMMPS dump file format 
\cite{LAMMPS}; if specified, a filter file is also read. Next, the \vpp{} library \cite{2009rycroft, 
lu2023extension} is used to compute the Voronoi cell of each particle, and \textit{VoroTop} 
computes the Voronoi cell topologies. Finally, the system is analyzed using features specified by the user and output is saved to disk; all output is saved in plain-text format.

\subsection{Command-line options}
\label{sec:options}
Features of the \textit{VoroTop} program are controlled through command-line options.
Some features described previously in Ref.~\cite{lazar2017vorotop} are omitted.

\subsubsection*{\texttt{-2}\hspace{4mm} two-dimensional system}
Interpret the data as describing a two-dimensional system. If $x$, $y$, and $z$ coordinates 
are all specified, then only the $x$ and $y$ coordinates are considered.

\subsubsection*{\texttt{-f}\hspace{4mm} load filter file}
Specifies a filter file to use for analysis. If this option is used, then a new LAMMPS dump 
output file will be created that includes the original data plus the structure types as 
determined by the given filter.

\subsubsection*{\texttt{-p}\hspace{4mm} $p$-vectors}
The Voronoi topology of each particle in the system is computed and saved to disk. The 
following information is recorded for the Voronoi cell of each particle: its number of edges, its 
number of neighbors with 3, 4, 5, etc.~edges, its canonical $p$-vector, the order of its 
symmetry group, and its chirality. Left-handed chirality is indicated by $-1$, right-handed 
chirality is indicated by 1, and a non-chiral Voronoi topology is indicated by 0.

\subsubsection*{\texttt{-d}\hspace{4mm} distribution of $p$-vectors}
This option calculates the distribution of Voronoi topologies in a system, and records it as a 
histogram of $p$-vectors.

\subsubsection*{\texttt{-c}\hspace{4mm} cluster analysis}
This feature implements the cluster analysis described in Section \ref{sec:clusteranalysis}. 
Each defect and crystal cluster is assigned a unique index, ordered by size. Positive indices 
indicate crystal clusters; negative indices indicate defect clusters. Also recorded for each 
particle is the size of the cluster to which it belongs. Particles with structure types listed in 
the specified filter are treated as crystalline, and defect clusters are built from particles 
whose structure types are not listed. 

\subsubsection*{\texttt{-r}\hspace{4mm} resolve indeterminate topologies}
This feature implements the analysis described in Section \ref{sec:indeterminate}; it is currently in testing form. Particles with indeterminate types are resolved by consideration of other particles in the same defect cluster. 

\subsubsection*{\texttt{-v}\hspace{4mm} Voronoi pair correlation function}
Computes the Voronoi pair correlation function for the system as described in 
Ref.~\cite{worlitzer2023pair}. This is the average number of Voronoi neighbors at each 
Voronoi distance from a central particle, averaged over all particles and normalized by data 
from the ideal gas. If an integer is specified, then the program computes the Voronoi pair 
correlation function up to that maximum Voronoi distance $k$; if left unspecified, data will be computed up to $k=50$. 

The \texttt{-u} option outputs the unnormalized version of the Voronoi pair correlation 
function. This is the average number of Voronoi neighbors at each Voronoi distance from a 
central particle, averaged over all particles.

\subsubsection*{\texttt{-e}\hspace{4mm} Encapsulated Postscript}
Outputs an encapsulated PostScript (eps) image of the system's particles and Voronoi cells. 
Table \ref{table:colors} lists different coloring scheme options. If no color scheme is specified 
then particles are colored according to the number of edges of their Voronoi cells. The 
\texttt{-n} flag can be added to specify that only the particles themselves be drawn, and not 
the Voronoi cells.
\begin{table}[h]
\begin{center}
\begin{tabular}{|c| l|}
 \hline
\textbf{Flag value} & \textbf{Particle coloring} \\
 \hline\hline
\texttt{0} & do not draw particles \\
 \hline
\texttt{1} & color all particles black \\
 \hline
\texttt{2} & color by number of edges \\
 \hline
\texttt{3} & color by filter index \\
 \hline
\texttt{4} & color by Voronoi distance from center \\
 \hline
\end{tabular}
\end{center}
\vspace{-3mm}
\caption{Color schemes for the \texttt{-e} option.
\vspace{3mm}
}
\label{table:colors}
\end{table}

Drawing all particles may be undesirable for large systems. If the \texttt{-e} flag is followed 
by two numbers, then the first specifies the coloring scheme and the second specifies the 
number of particles that should be drawn; a window centered at the middle of the system 
and whose area is proportional to the number of particles specified is drawn.

\subsection{Limitations}
\label{sec:limitations}

At present, \textit{VoroTop} cannot distinguish between particles of different sizes or 
chemical types. A future version of \textit{VoroTop} will handle particles of different sizes 
using the radical Voronoi tessellation, a generalization of the standard Voronoi tessellation; 
computation of the radical Voronoi tessellation is already available in \vpp. Analysis of 
multicomponent systems will require generalizing the canonical representation introduced in 
Section \ref{sec:canonical}; implementation of this analysis in \textit{VoroTop} will require 
new data structures and algorithms.

\section{Application examples}
\label{sec:applications}
\subsection{Identifying defects in polycrystalline systems}
\label{sec:sub:identifying}

\begin{figure}
\begin{center}
\fbox{\begin{overpic}[width=0.49\columnwidth]{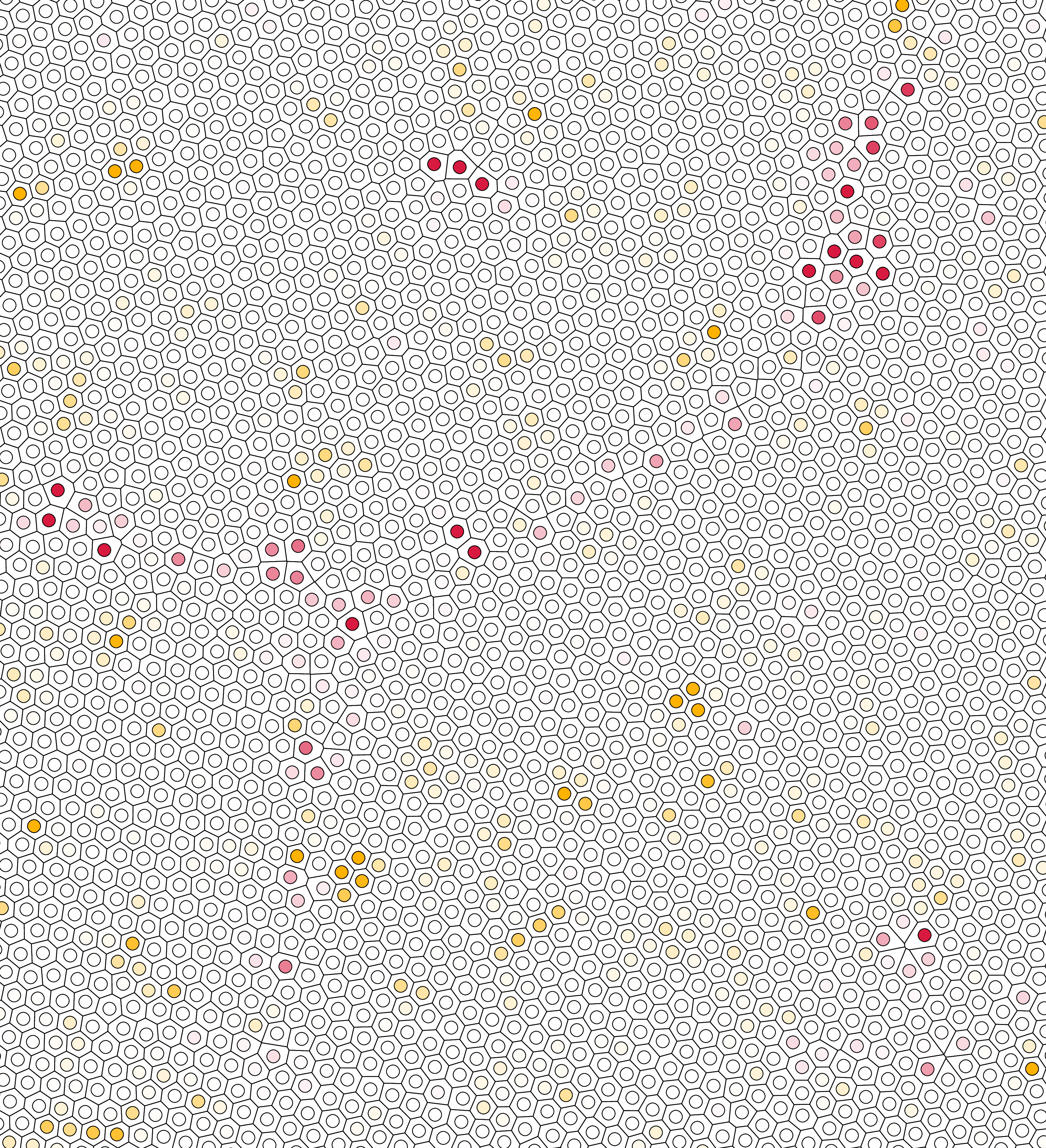}
\put (0,930.) {\setlength{\fboxsep}{1pt} {\fcolorbox{black}{white}{(a)}}}
\end{overpic}} \hfill
\fbox{\begin{overpic}[width=0.49\columnwidth]{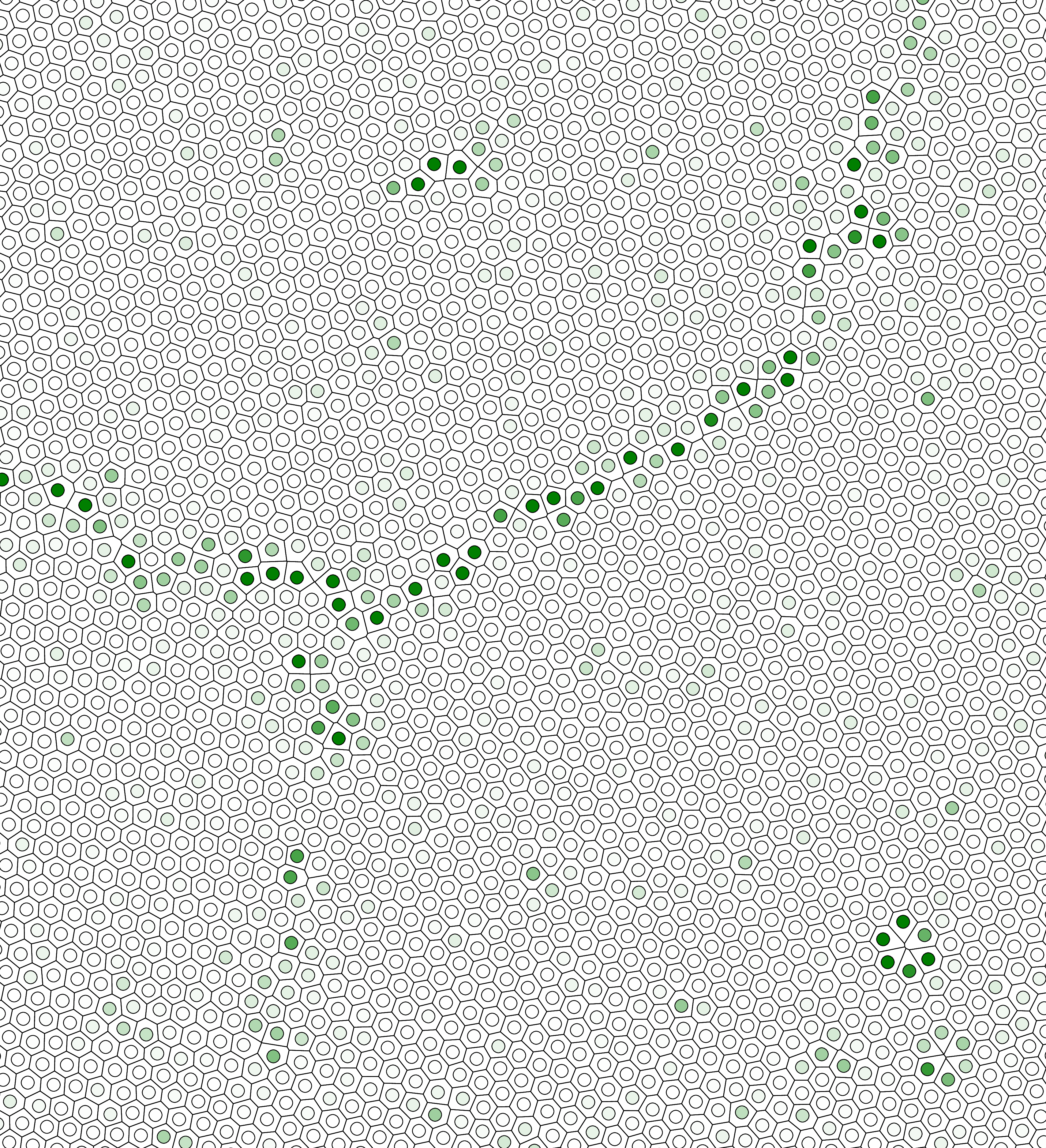}
\put (0,930.) {\setlength{\fboxsep}{1pt} {\fcolorbox{black}{white}{(b)}}}
\end{overpic}} \vspace{-2.5mm}\\
\fbox{\begin{overpic}[width=0.49\columnwidth]{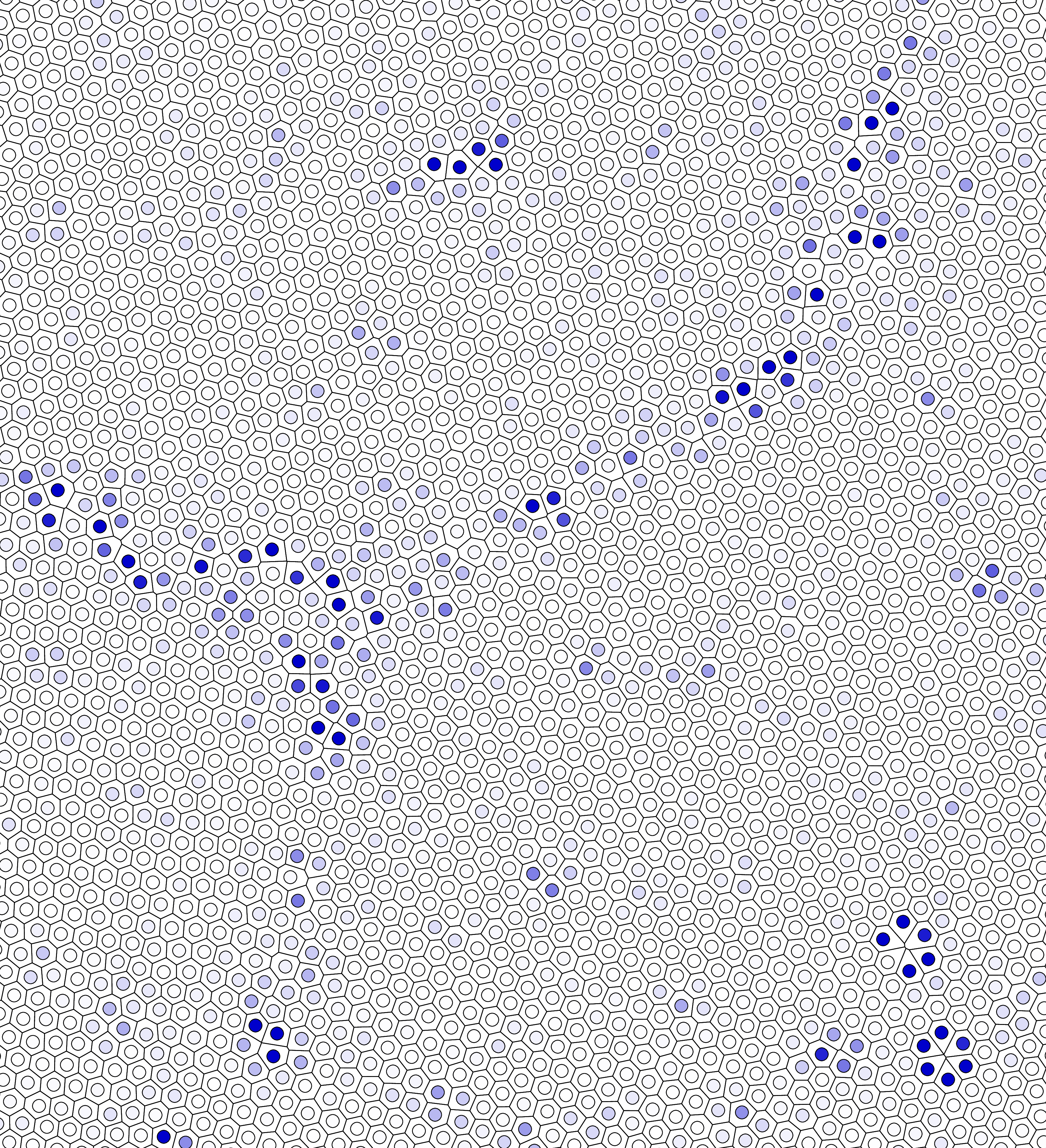}
\put (0,930.) {\setlength{\fboxsep}{1pt} {\fcolorbox{black}{white}{(c)}}}
\end{overpic}} \hfill
\fbox{\begin{overpic}[width=0.49\columnwidth]{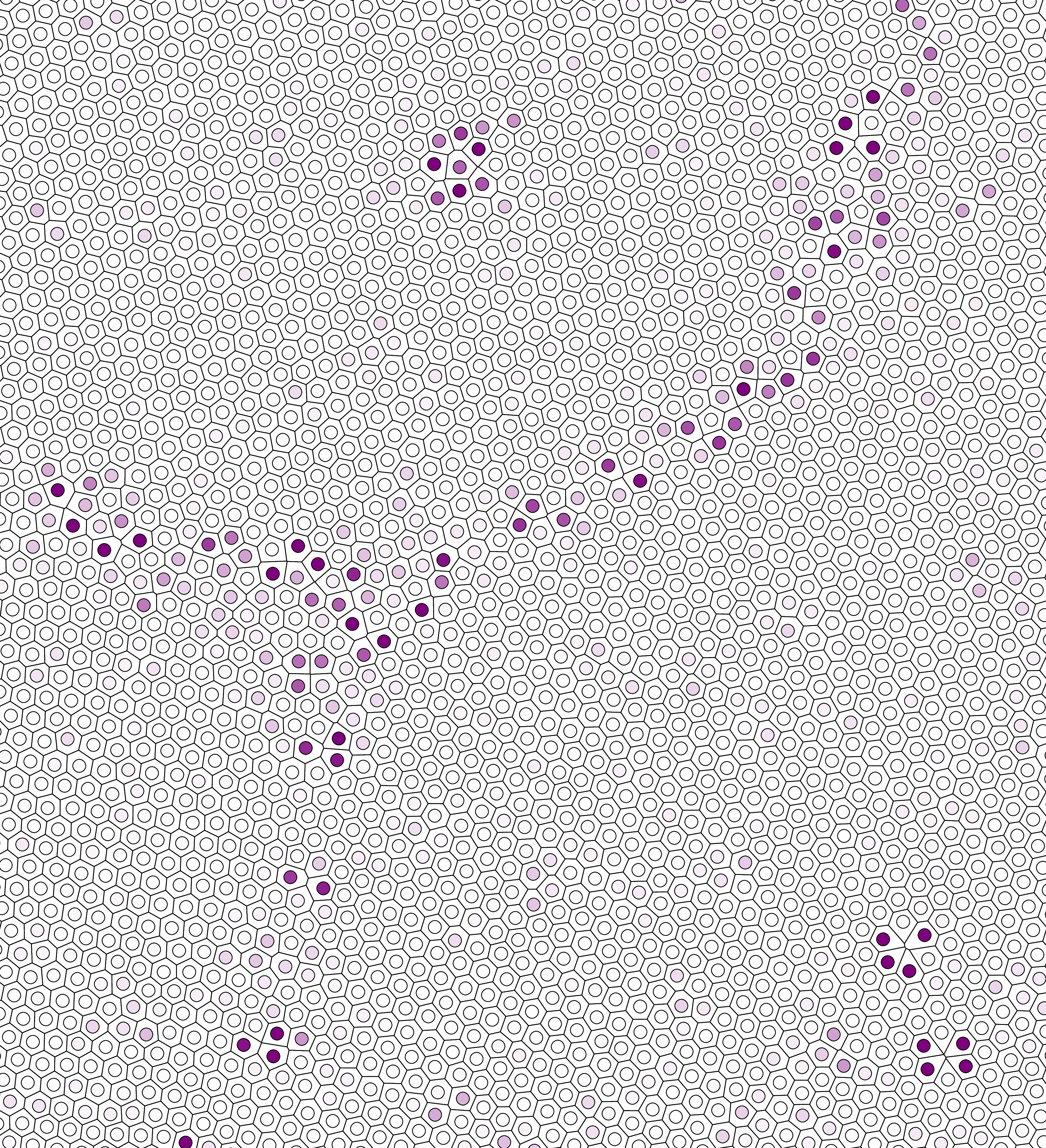}
\put (0,930.) {\setlength{\fboxsep}{1pt} {\fcolorbox{black}{white}{(d)}}}
\end{overpic}}
\end{center}
\caption{A polycrystal with vacancies, dislocations, and grain boundaries created using 
molecular dynamics through the cooling of a Lennard-Jones liquid. Particles colored 
according to (a) Voronoi cells areas, with larger ones colored red, and smaller ones colored 
yellow;
(b) centrosymmetry,
(c) bond-angle analysis, and
(d) the variance in distances to Voronoi neighbors.}
\label{fig:two-dimensional-othermethods}
\end{figure}

\begin{figure}
\begin{center}
\fbox{\begin{overpic}[width=0.49\columnwidth]{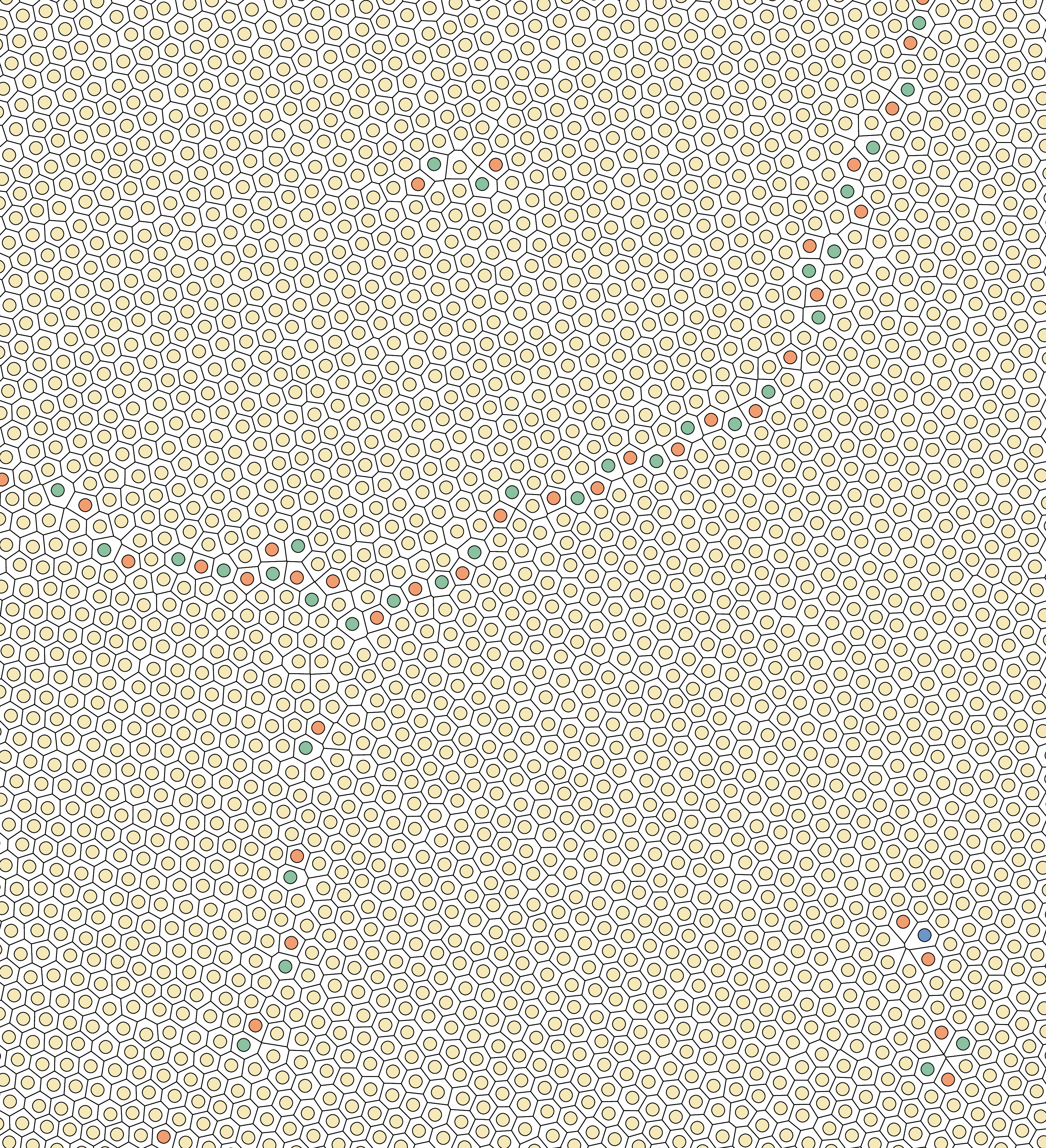}
\put (0,930.) {\setlength{\fboxsep}{1pt} {\fcolorbox{black}{white}{(a)}}}
\put (400,10) {\includegraphics[width=0.12\columnwidth]{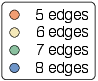}}
\end{overpic}} \hfill
\fbox{\begin{overpic}[width=0.49\columnwidth]{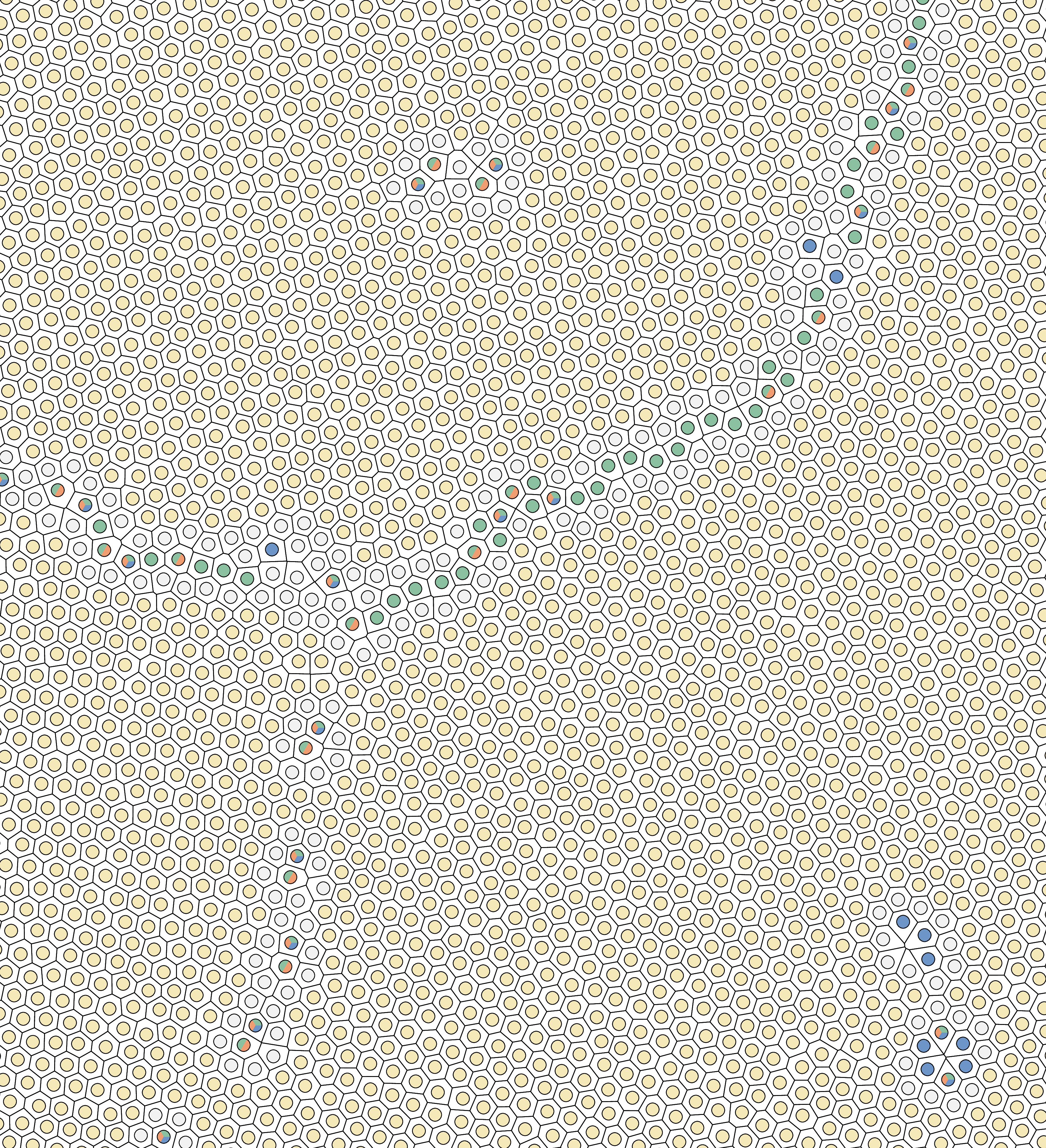}
\put (0,930.) {\setlength{\fboxsep}{1pt} {\fcolorbox{black}{white}{(b)}}}
\put (360,10) {\includegraphics[width=0.18\columnwidth]{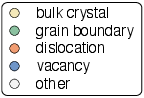}}
\end{overpic}} \vspace{-2.5mm}\\
\fbox{\begin{overpic}[width=0.998\columnwidth]{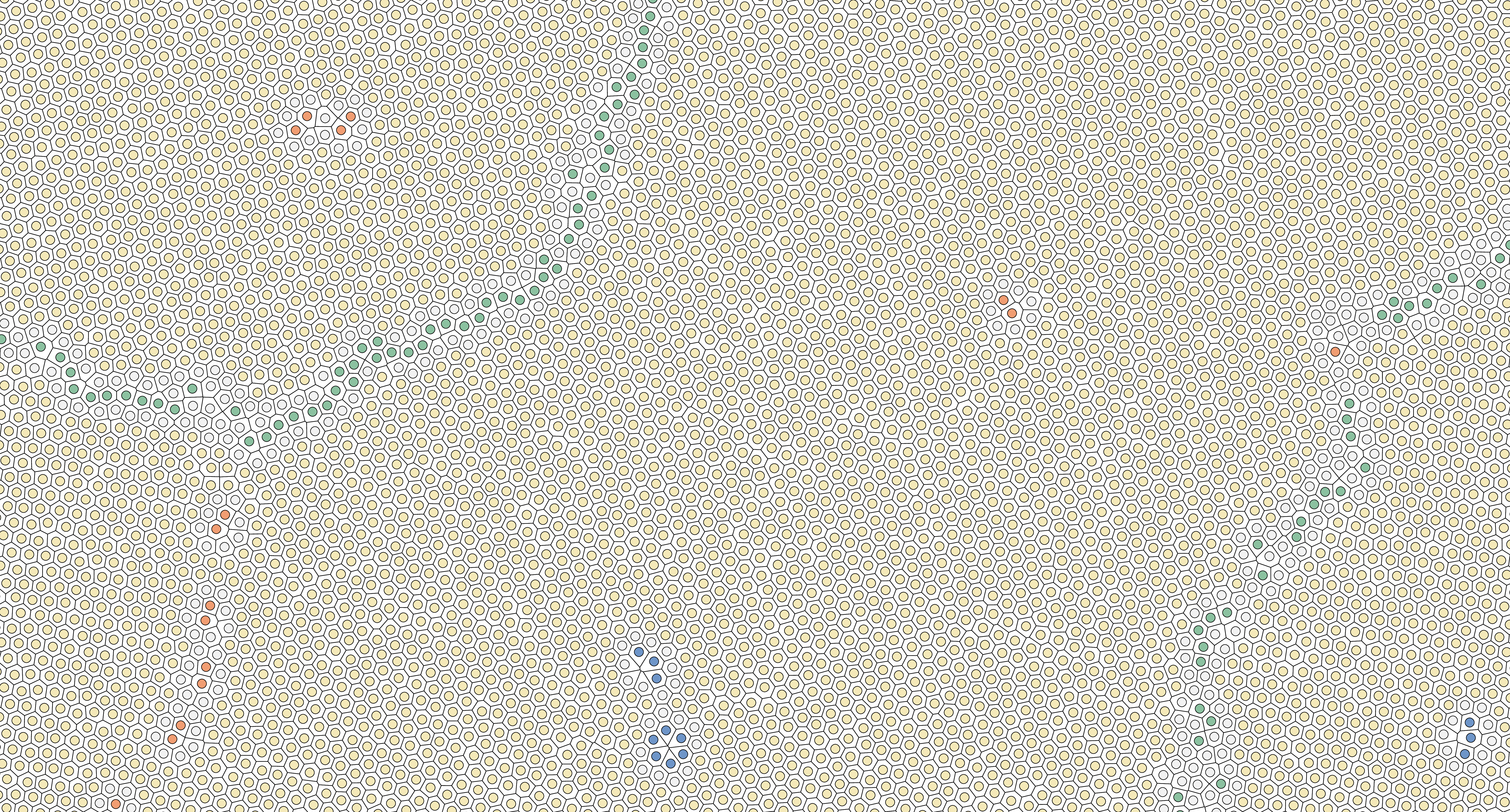}
\put (0,500.) {\setlength{\fboxsep}{1pt} {\fcolorbox{black}{white}{(c)}}}
\put (190,5) {\includegraphics[width=0.18\columnwidth]{key7b.eps}}
\end{overpic}} 
\end{center}
\caption{A polycrystal with vacancies, dislocations, and grain boundaries created using 
molecular dynamics through the cooling of a Lennard-Jones liquid. Particles colored 
according to (a) the number of edges of each particle; (b) Voronoi topology using a simple 
filter; indeterminate types are colored using multiple colors; (c) Voronoi topology after 
indeterminate types are resolved using cluster analysis.}
\label{fig:two-dimensional-vt-methods}
\end{figure}

To illustrate the effectiveness of Voronoi topology in characterizing and visualizing defects in 
crystalline systems, Figures \ref{fig:two-dimensional-othermethods} and 
\ref{fig:two-dimensional-vt-methods} illustrate part of a two-dimensional polycrystal using several standard methods, and with Voronoi topology. 
The simulated system was constructed using a molecular dynamics simulation of a Lennard-Jones liquid, which was cooled until it crystalized and then annealed at half of its 
bulk melting temperature.  Although it is 
possible to identify visually the rough contours of grain boundaries, dislocations, and 
vacancies from the particles themselves, automating further analysis requires an algorithmic 
approach.

\textbf{Standard methods.}
Particles in Figure \ref{fig:two-dimensional-othermethods}(a) are colored according to the 
areas of their Voronoi cells. Particles whose Voronoi cells have smaller than average areas 
are colored yellow, while those with larger than average areas are colored red. Figure 
\ref{fig:two-dimensional-othermethods}(b) shows the same system but with particles colored 
according to centrosymmetry \cite{Kelchner_centrosymmetry}; darker shades indicate higher 
values. Particles in Figure \ref{fig:two-dimensional-othermethods}(c) are colored according 
to a bond-angle order parameter, in particular the sample variation of the angles formed by 
adjacent pairs of Voronoi neighbors. Finally, the particles in Figure 
\ref{fig:two-dimensional-othermethods}(d) are colored according to the sample variation of 
the distances to Voronoi neighbors. Generally speaking, particles belonging to defects have 
order parameter values that are different from those associated to particles belonging to bulk 
crystals. Thus, with each approach, defects can be detected through the presence of 
particles colored in darker shades.

Classifying particles as belonging to either a bulk crystal or else to a defect requires 
choosing an order-parameter cutoff. At low temperatures, there exists a gap between 
order-parameter values associated with the bulk crystal and those associated with defects. 
Any choice of cutoff in that gap will thus result in the same binary classification of particles. 
At finite temperatures, however, and especially at high temperatures, thermal fluctuations 
result in bulk crystal particles that have order parameter values associated with defects. 
Consequently, any choice of order parameter cutoff will result in bulk crystal particles that 
are misidentified as belonging to defects, defect particles misidentified as belonging to a bulk 
crystal, or both. Consequently, the particle-level details of defects are often difficult to 
discern.

Moreover, even when conventional order parameters can reliably detect the presence of a 
defect, they typically cannot distinguish between defects of different kinds. Order-parameter 
values associated with a vacancy, for example, might coincide with those associated with a 
dislocation or grain boundary. Thus, even at low temperatures, when the distinction between 
locally crystalline particles and those associated to defects is clear, distinguishing different 
kinds of defects is still challenging.

\textbf{Voronoi topology.}
Figure \ref{fig:two-dimensional-vt-methods} uses Voronoi topology to characterize and 
visualize individual particles. Particles in Figure \ref{fig:two-dimensional-vt-methods}(a) are 
colored according to the number of edges of their Voronoi cells. This basic approach 
highlights the efficacy of using topological features of the Voronoi cells --- particles belonging to crystals have hexagonal Voronoi cells, while those belonging to structural defects have Voronoi cells with other numbers of edges.

Particles in Figure \ref{fig:two-dimensional-vt-methods}(b) are colored using a filter of 
Voronoi topologies associated with bulk crystals, grain boundaries, dislocations, and 
vacancies.  The list of Voronoi topologies, denoted by $p$-vectors, used to color this figure can be found in Table 
\ref{table:filter}. Note that certain topologies are associated with multiple defects, and are 
hence called indeterminate and colored multiple colors, corresponding to their multiple associated structures.
\begin{table}
\centering
\begin{tabular}{ | l  l |}
\hline
\multicolumn{2}{| l | }{{\bf Crystal }}\\
& (6,6,6,6,6,6,6) \\
&  \\
\multicolumn{2}{| l | }{{\bf Grain boundary }}\\
& (5,6,6,6,6,7) \\
& (5,6,6,7,6,7) \\
& (6,5,6,6,6,7,6) \\
& (6,5,6,6,7,6,6) \\
& (7,5,6,6,5,6,6,6) \\
& (7,5,6,6,6,6,6,6) \\
\hline
\end{tabular}
\hspace{5mm}
\begin{tabular}{ | l  l |}
\hline
\multicolumn{2}{| l | }{{\bf Vacancy }}\\
& (5,6,6,6,6,7) \\
& (5,6,6,6,6,8) \\
& (5,6,6,6,7,7) \\
& (6,5,6,6,6,6,8)\\
& (6,5,6,6,6,7,7) \\
& (6,6,6,6,6,6,8)\\
& (7,5,6,6,6,5,7,7) \\
& (7,5,6,6,6,6,7,6) \\
& (8,5,6,6,6,5,6,6,6) \\
\hline
\end{tabular}
\hspace{5mm}
\begin{tabular}{ | l  l |}
\hline
\multicolumn{2}{| l | }{{\bf Dislocation }}\\
& (5,6,6,6,6,7) \\
& (7,5,6,6,6,6,6,6)\\
& \\
\multicolumn{2}{| l | }{{\bf Interstitial }}\\
& (5,6,6,7,6,7) \\
& (6,5,7,5,7,5,7) \\
& (7,5,6,5,6,6,6,6)\\
&  \\
&  \\
\hline
\end{tabular}
\caption{A list of structural defects and associated Voronoi topologies, denoted by their $p$-vectors. Note that some topologies are associated with multiple defects.}
\label{table:filter}
\end{table}
Particles with other Voronoi 
topologies are colored light grey. This visualization provides a clear picture of the bulk 
crystals as well as vacancies, interstitials, and grain boundaries.

Finally, Figure \ref{fig:two-dimensional-vt-methods}(c) shows the result of a 
post-processing cluster analysis to resolve indeterminate types and to identify structural 
defects such as dislocations, vacancies, and grain boundaries. In particular, we considered 
contiguous sets of particles with non-crystalline Voronoi topologies. A cluster containing one 
of each of the topologies associated with a dislocation, and none of the other topologies in 
Table \ref{table:filter}, is identified as a dislocation.  Defect clusters with exactly three or six 
particles with topologies associated with vacancies are identified as vacancies.  Finally, 
contiguous sets of particles all of whose topologies are associated with grain boundaries are 
identified as grain boundaries.

\subsection{Characterizing order in disordered systems}

Voronoi topology analysis can also be used to characterize and analyze nominally 
disordered systems. In crystalline systems, local arrangements of particles are all of the 
same kind, or else of a small number of kinds. This order is reflected in the relatively small 
number of Voronoi topologies observed in such systems. In the hexagonal crystal, for 
example, all Voronoi cells are hexagons and have the $p$-vector $(6,6,6,6,6,6,6)$, even 
after particle coordinates are perturbed. Systems with multiple particles in a repeating unit 
cell may be associated with several Voronoi topologies, though this number is always finite.

In contrast, nominally disordered systems can have an infinite number of possible 
arrangements of particles due to their lack of long-range periodic order. A statistical 
description of the relative frequencies of distinct particle arrangements, as classified through 
Voronoi topology, is one way to describe local structural features in these systems. We thus 
consider the distribution of Voronoi topologies in several disordered systems.

We consider three examples: an ideal gas, a Lennard-Jones liquid heated to 150\% of its 
bulk melting temperature, and a hyperuniform system constructed using a Vicsek model of 
collective motion \cite{vicsek1995novel}. To sample the ideal gas, we generated 80 systems, 
each containing 4 million points, randomly distributed in the unit square with periodic 
boundary conditions. To sample from the Lennard-Jones liquid, we constructed 1600 
systems, each containing 17,280 particles. We used a Vicsek model with one million 
particles, unit density, and uniform noise in $[-0.6\pi ,0.6\pi ]$; simulations were run for 
50,000 time steps.
Particles in the three systems are illustrated in Figure \ref{fig:systems-and-dist-p}.
\setlength{\fboxsep}{0.25pt}
\begin{figure}
\centering
\fbox{\begin{overpic}[width=0.3\columnwidth]{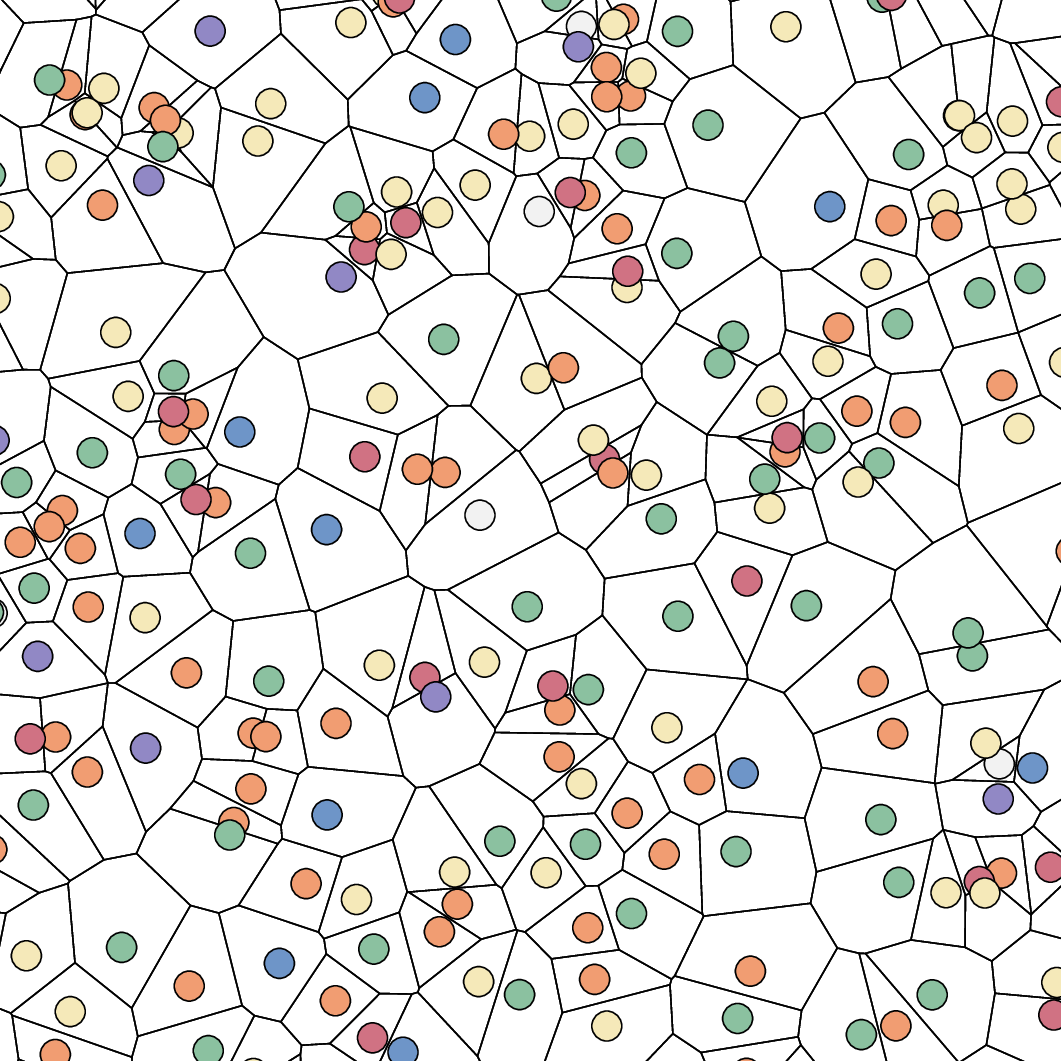}
\put (-10.,875) {\setlength{\fboxsep}{1pt} {\large \fcolorbox{black}{white}{(a)}}}
\end{overpic}}
\hfill
\fbox{\begin{overpic}[width=0.3\columnwidth]{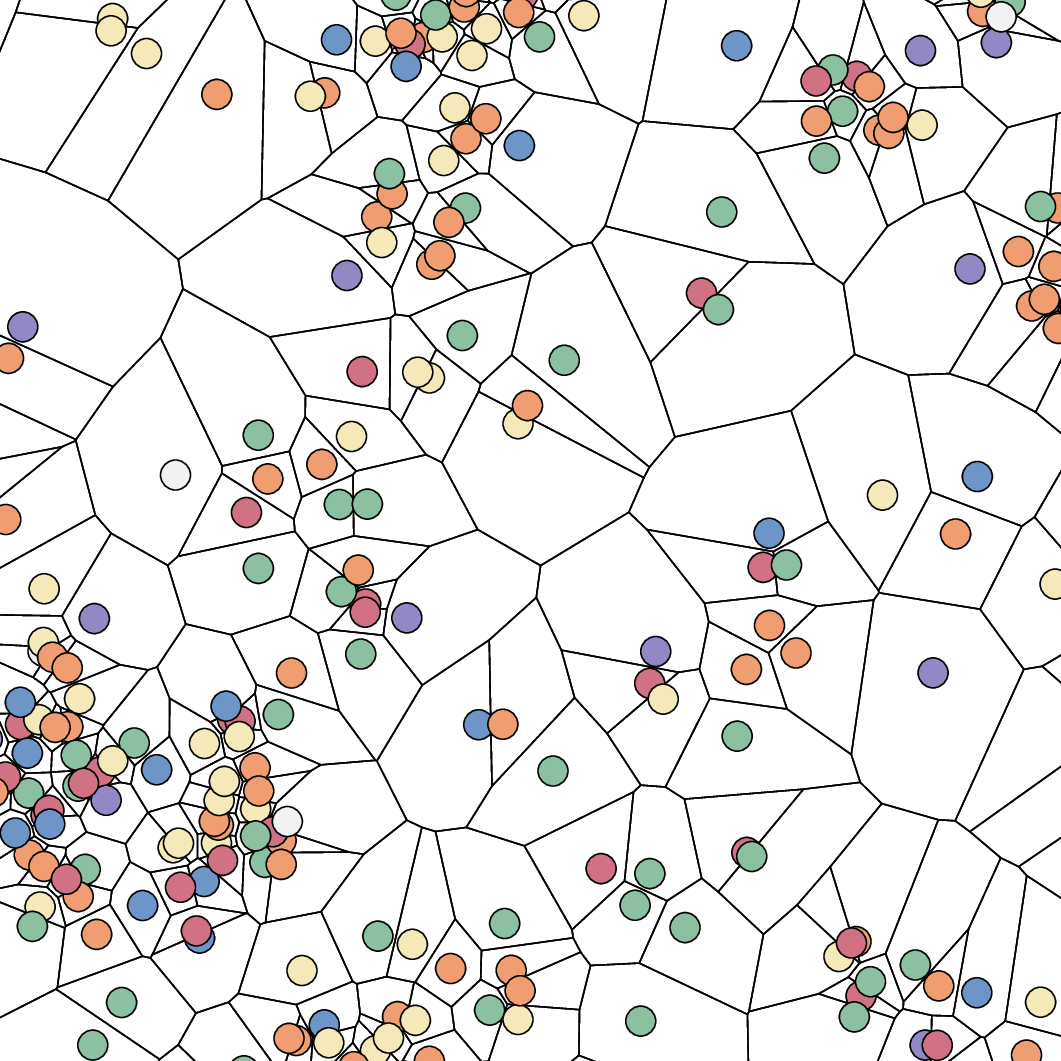}
\put (-10.,875) {\setlength{\fboxsep}{1pt} {\large \fcolorbox{black}{white}{(b)}}}
\end{overpic}}
\hfill
\fbox{\begin{overpic}[width=0.3\columnwidth]{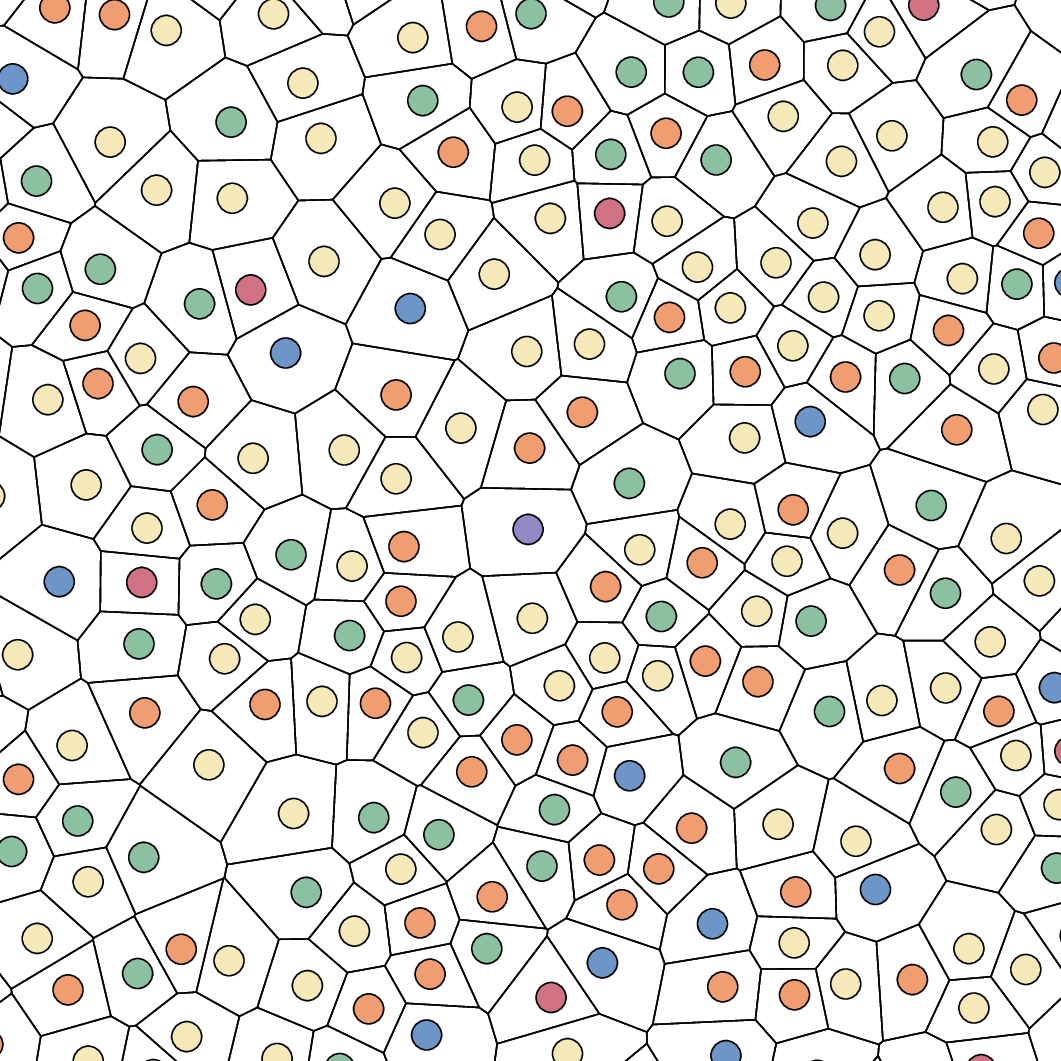}
\put (-10.,875) {\setlength{\fboxsep}{1pt} {\large \fcolorbox{black}{white}{(c)}}}
\end{overpic}}
\hfill 
\includegraphics[height=0.30\columnwidth]{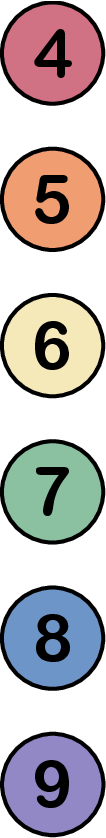}
\caption{Images of particles in (a) an ideal gas, (b) a Vicsek model of collective 
motion, and (c) a Lennard-Jones liquid heated to 150\% of its bulk melting temperature;
each figure has roughly 200 particles.
Colors indicate numbers of Voronoi edges or neighbors.}
\label{fig:systems-and-dist-p}
\end{figure}

Table \ref{table:sides} shows the frequencies of particles with different numbers of Voronoi cell edges 
or neighbors in the three systems.
Although the average number of Voronoi cell edges must be six in all of them, the distribution of number of 
edges differs.  Notice in particular that this distribution appears narrowest in the Lennard-Jones liquid, 
reflecting what appears to be a more regular kind of disorder as compared with that in the other systems. 
\begin{table}
\centering
\begin{tabular}{ | r | wr{1.65cm} | wr{1.65cm} | wr{1.65cm} |}
\hline
\multicolumn{4}{| c | }{{\bf Distribution of Voronoi cell sides}}\\
\hline
\multicolumn{1}{| c | }{{\bf Sides }} & \multicolumn{1}{| c | }{{\bf Ideal gas }} & \multicolumn{1}{| c | }{{\bf Vicsek }} & \multicolumn{1}{| c | }{{\bf LJ liquid }} \\
\hline
3 & 1.12 & 1.47 & 0.05 \\
4 & 10.69 & 12.11 & 4.94  \\
5 & 25.94 & 26.12 & 27.69 \\
6 & 29.47 & 27.48 & 38.42 \\
7 & 19.88 & 18.42 & 21.46 \\
8 & 9.01 & 9.10 & 6.22 \\
9 & 2.97 & 3.63 & 1.09 \\
10 & 0.74 & 1.20 & 0.13 \\
11 & 0.15 & 0.35 & 0.01 \\
12 & 0.02 & 0.08 & 0.00 \\
\hline
\end{tabular}
\caption{The fraction of particles (\%) in each system with a given number of Voronoi 
sides or neighbors in the ideal gas, a Vicsek model of collective motion, and a 
Lennard-Jones liquid.}
\label{table:sides}
\end{table}

Table \ref{table:pvectortable} tabulates the frequencies of the most common Voronoi 
topologies observed in these systems. Despite their different origins, the ideal gas 
and Vicsek model system appear most structurally similar, judging 
by frequencies of Voronoi topologies in the systems. In contrast, the distribution of Voronoi 
topologies appears qualitatively different in the Lennard-Jones liquid. In particular, the most common types in 
the Lennard-Jones liquid appear roughly three times more frequently than the most common 
types in the other systems. These differences in particle arrangements in the different 
systems likely reflect different energetic and entropic forces that govern their behavior.

In all systems, the relatively high frequencies of Voronoi topologies whose central particle 
have only four or five edges might appear puzzling, given that six-sided Voronoi cells are 
the most common in all systems.  This can be understood as a combinatorial result of the 
increasing number of possible Voronoi topologies as the number of neighbors of a central 
particle increases.  Since there are many more ways of arranging 6 neighbors, for example, 
than only 5, the relative frequency of many arrangements with five neighbors will be greater 
than those with 6.

\begin{table}
\centering
\begin{tabular}{ | l | c |}
\hline
\multicolumn{2}{| c | }{{\bf Ideal gas }}\\
\hline
{\bf $p$-vector} & $f(\%)$ \\
\hline
$(4,6,6,7,8)	$ & 0.3502\\
$(5,5,6,7,6,7)	$ & 0.3372\\
$(5,5,6,6,6,7)	$ & 0.3371\\
$(5,5,6,6,7,7)	$ & 0.3141\\
$(5,5,6,6,7,6)	$ & 0.3123\\
$(5,5,7,6,7,7)	$ & 0.3110\\
$(4,5,6,7,7)	$ & 0.3000\\
$(4,6,7,7,8)	$ & 0.2978\\
$(4,6,6,7,7)	$ & 0.2954\\
$(4,6,6,6,7)	$ & 0.2887\\
\hline
\end{tabular}
\hfill
\begin{tabular}{ | l | c |}
\hline
\multicolumn{2}{| c | }{{\bf Vicsek model }}\\
\hline
{\bf $p$-vector} & $f(\%)$ \\
\hline
$(4,5,6,6,7)$ & 0.3691 \\
$(4,5,6,7,7)$ & 0.3480 \\
$(5,5,6,6,6,7)$ & 0.3276 \\
$(4,6,6,6,7)$ & 0.3095 \\
$(5,5,6,6,7,6)$ & 0.3091 \\
$(4,6,6,7,8)$ & 0.2921 \\
$(4,5,7,6,8)$ & 0.2761 \\
$(4,5,6,6,8)$ & 0.2732 \\
$(5,5,6,7,6,7)$ & 0.2659 \\
$(4,6,6,7,7)$ & 0.2586 \\
\hline
\end{tabular}
\hfill
\begin{tabular}{ | l | c |}
\hline
\multicolumn{2}{| c | }{{\bf Lennard-Jones liquid }}\\
\hline
{\bf $p$-vector} & $f(\%)$ \\
\hline
$(5,6,6,6,6,7)$ & 1.1139\\
$(6,5,6,6,6,6,7)$ & 0.9264\\
$(5,6,6,7,6,7)$ & 0.9088\\
$(5,5,6,7,6,7)$ & 0.8961\\
$(5,5,6,6,6,7)$ & 0.8342\\
$(5,5,7,6,7,7)$ & 0.8190\\
$(6,5,6,6,6,7,6)$ & 0.7879\\
$(5,6,6,6,7,7)$ & 0.7838\\
$(5,5,6,6,7,7)$ & 0.7608\\
$(5,5,6,6,7,6)$ & 0.6780\\
\hline
\end{tabular}
\caption{Lists of the ten most common $p$-vectors and their frequencies \emph{f} in three
nominally disordered systems: the ideal gas, a Vicsek model, and a Lennard-Jones liquid 
heated to 150\% of its bulk melting temperature.}
\label{table:pvectortable}
\end{table}

\subsection{Characterizing real grain boundaries}

\begin{figure}[h]
\begin{minipage}[b]{.5\textwidth}
\subfloat {\fbox{\begin{overpic}[trim={230cm 7cm 230cm 10cm},clip,width=1.0\linewidth]{figure9a.eps} \put (-10,190) {\setlength{\fboxsep}{1pt} {\fcolorbox{black}{white}{(a)}}} \end{overpic}}}\\ 
\subfloat {\fbox{\begin{overpic}[trim={230cm 7cm 230cm 10cm},clip,width=1.0\linewidth]{figure9b.eps}\put (-10,190) {\setlength{\fboxsep}{1pt} {\fcolorbox{black}{white}{(b)}}} \end{overpic}}}
\vspace{3mm}
\end{minipage}
\hfill
\begin{minipage}[b]{.45\textwidth}
\subfloat {
\begin{overpic}[width=\textwidth]{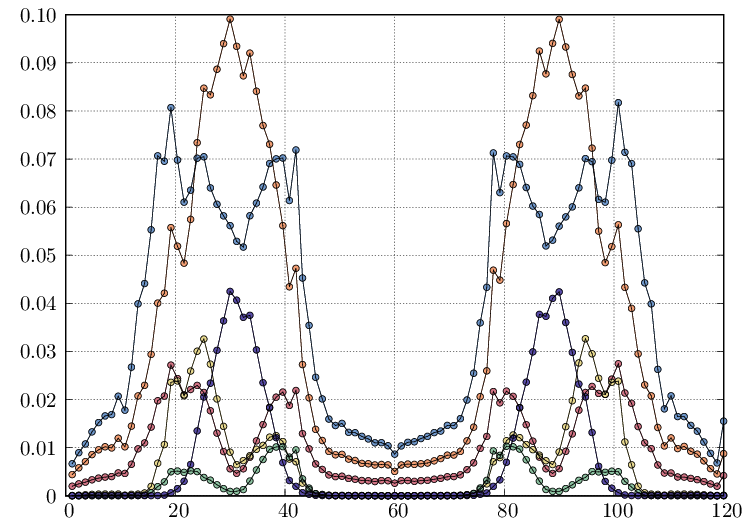}
\put (75,615.) {\setlength{\fboxsep}{1pt} {\fcolorbox{black}{white}{(c)}}}
\put (200.,-45) {\footnotesize Misorientation angle (deg)}
\put (415,495.) {\includegraphics[width=0.23\columnwidth]{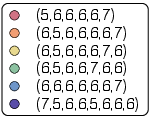}}
\put (-40.,25) {\footnotesize \rotatebox{90}{Number density per length}}
\end{overpic}}
\end{minipage}
\vspace{2mm}
\caption{(a) High- and (b) low-angle real symmetric tilt grain boundaries in two-dimensional 
bicrystals; particles colored according to the number of edges of their Voronoi cells. (c) The 
number density of various Voronoi topologies per unit length as a function of misorientation
angle; each color indicates a different Voronoi topology.}
\label{fig:bicrystals}
\end{figure}

A significant challenge that arises in studying grain boundaries in realistic systems is their 
structural complexity as compared with grain boundaries in ideal systems. Thus, while 
perfect symmetric tilt grain boundaries, for example, can be described in the language of 
bicrystallography \cite{pond1983bicrystallography} and structural unit models 
\cite{bishop1968coincidence, balluffi1984structural, phillpot1995structural, han2017grain}, 
those in realistic ones typically cannot. Figures \ref{fig:bicrystals}(a) and (b) illustrate high- 
and low-angle symmetric tilt grain boundaries in a two-dimensional, Lennard Jones bicrystal 
annealed at half of its bulk melting temperature. The irregular nature of these grain 
boundaries complicate their description.

Voronoi topology can be used to provide a statistical characterization of order in local 
structural terms. Figure \ref{fig:bicrystals}(c) illustrates the frequencies of different Voronoi 
topologies in realistic symmetric tilt grain boundaries as a function of misorientation angle. 
This approach provides a robust characterization of grain boundary structure that is largely 
independent of microdegrees of freedom \cite{sutton2015five}. Moreover, this 
characterization can be useful in solving a related inverse problem -- given a set of particle 
positions can we determine the misorientation angle? Figure \ref{fig:bicrystals}(c) suggests 
that knowledge of the distribution of Voronoi topologies, or even just the relative frequencies 
of several common types, is sufficient to identify the misorientation between the two grains. 
Voronoi topology thus provides a method to robustly characterize complex structure in 
statistical-structural terms. Analysis of energetic features of particle arrangements might 
provide insight into energetic aspects of realistic grain boundaries.

\subsection{Chirality in grain boundaries}
\label{sec:chiral}

\begin{figure}
\fbox{\begin{overpic}[width=\columnwidth]{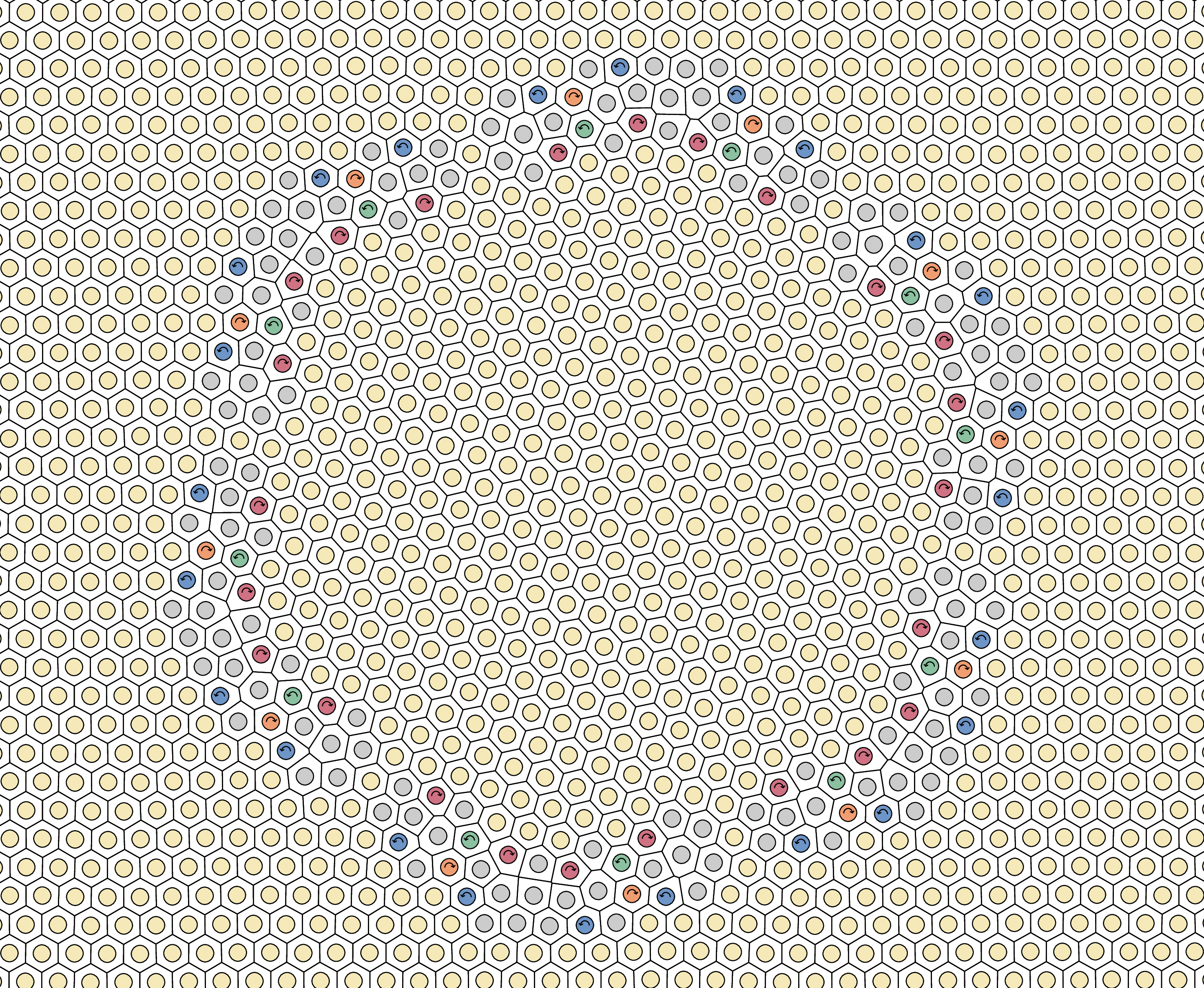}
\put (745,660.) {\includegraphics[width=0.25\columnwidth]{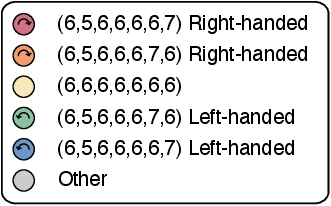}}
\end{overpic}}
\caption{A circular grain boundary in a two-dimensional hexagonal Lennard-Jones bicrystal. 
The inside grain was constructed by rotating a circular region by $16^{\circ}$, and then 
annealing the system at 50\% of its bulk melting temperature. The orientation of the grain 
boundary can be observed in Voronoi topologies of the particles.}
\label{two-dimensional-circular}
\end{figure}

A unique strength of the present approach towards structure characterization is its ability to 
identify chiral features of particle arrangements. As described in Section \ref{sec:canonical}, 
certain arrangements of particles lack a mirror symmetry, and hence can be distinguished 
from their mirror images. Although the canonical representation of Voronoi topology via the 
$p$-vector ignores differences in orientations, this information is recorded while computing 
the $p$-vector.

Figure \ref{two-dimensional-circular} illustrates a circular grain boundary in a 
two-dimensional hexagonal Lennard-Jones bicrystal heated to 50\% of its bulk melting 
temperature. For some misorientation angles, we expect that right-handed and left-handed 
versions of particle arrangements, as classified through Voronoi topology, appear in the 
same proportions. Even if thermal vibrations result in local differences, these differences 
should be negligible for large samples. However, for other misorientation angles, grain 
boundaries can exhibit an orientation, even in two dimensions. This can be examined 
through Voronoi topology analysis.

Particles in Figure \ref{two-dimensional-circular} are colored according to their Voronoi 
topology and orientation, as indicated in the key. In addition, particles with oriented 
$p$-vectors are further labeled with directed arrows, to indicate whether they are right- or 
left-handed. Notice that all right-handed forms of the $p$-vector $(6,5,6,6,6,6,7)$, colored 
red, and all left-handed forms of $(6,5,6,6,6,7,6)$, colored green, appear on the inside part of 
the circular grain boundary, whereas particles with identical topologies but opposite 
orientations appear on the outside of the grain boundary.  
The appearance of chiral 
features on the single-particle scale results from a chirality of the grain boundary itself. 
Automating the analysis of chiral features in particle systems might aid in the study of grain 
rotation and its impact on grain growth in two-dimensional polycrystals 
\cite{hutchinson2024grain}.

\section{Discussion}
\label{sec:conclusions}

Voronoi topology provides an effective approach to characterizing structural features of two-dimensional particle systems.  As a topological method, it is generally insensitive to small perturbations of particle coordinates, making it particularly useful for analyzing imperfect systems, including finite-temperature crystals, and systems otherwise perturbed from their ground states.  Similarly, it is effective for analyzing experimental data, which is often characterized by some measurement error.  This robustness in the face of uncertainty is consistent with the intuition that structural features of particle systems do not change under small local perturbations.  

The effectiveness of the proposed approach in a broad range of applications -- identifying crystals and defects in high-temperature systems, characterizing disordered systems, non-ideal grain boundaries, and even chiral features of particle systems -- highlights its general utility.  Any one of these tasks can be challenging, and the ability to approach all of them with a single set of tools is noteworthy.

The proposed method is limited in certain respects.  Voronoi topology is naturally insensitive 
to questions of scale, and also cannot capture local density fluctuations.  To some degree, 
these limitations could be remedied by consideration of Voronoi cell areas and perimeters, or 
other geometric features of the particle positions.  The development of hybrid methods, 
integrating Voronoi topology with geometric information, might provide a more powerful 
approach with more general applications \cite{kaliman2016limits}.

Another current limitation of Voronoi topology as described above is its identical treatment of 
all particles.  The methods described above, as well as the $p$-vector notation, however, 
can be extended so to generalize the analysis for multicomponent systems such as those 
consisting of particles of different chemical types.  We leave these extensions to future 
work.

\section*{Acknowledgments}
This research was supported by a grant from the United States -- Israel Binational Science 
Foundation (BSF), Jerusalem, Israel through grant number 2018/170. Additional support of 
the Data Science Institute at Bar-Ilan University is also gratefully acknowledged. 
C.~H.~R.~was partially supported by the Applied Mathematics Program of the U.S.~DOE 
Office of Science Advanced Scientific Computing Research under Contract No.~DE-
AC02-05CH11231.

\bibliographystyle{ieeetr}
\bibliography{refs}

\end{document}